\documentclass[aip,amsmath,preprint,tightenlines]{revtex4-2}
\usepackage{dcolumn} 
\usepackage{bm} 
\usepackage[table]{xcolor}
\usepackage[bottom]{footmisc}
\usepackage[nolist]{acronym}
\usepackage{amsmath}
\usepackage{amssymb}    
\usepackage{graphicx}
\usepackage{tabularx}
\usepackage{booktabs}
\usepackage{multirow}
\usepackage{makecell}
\usepackage{rotating}

\usepackage[flushleft]{threeparttable}
\draft 
\usepackage{url} 
\newcommand\longvar[1]{\mathchardef\UrlBreakPenalty=100
\mathchardef\UrlBigBreakPenalty=100\url{#1}}
\usepackage{hyperref}
\usepackage[newfloat=true,frozencache]{minted}
\usepackage{pifont}
\usepackage{adjustbox}
\newcommand\mathplus{+}

\newcommand{\analog}[1]{\breve{#1}}
\newcommand{\pulsed}{\xrightarrow{\cdots}}
\newcommand{\rowpulsed}{\xrightarrow{.}}
\newcommand{\choppulsed}{\xrightarrow[+-]{\cdots}}

\newcommand{\figref}[1]{Fig.~\ref{fig:#1}}
\newcommand{\tabref}[1]{Tab.~\ref{tab:#1}}
\newcommand{\secref}[1]{Sec.~\ref{sec:#1}}
\renewcommand{\eqref}[1]{Eq.~\ref{eq:#1}}
\newcommand{\RPUConfig}{\textit{RPUConfig}}
\newcommand{\pytorch}{\textsc{pyTorch}}

\DeclareMathOperator{\quantize}{quant}
\DeclareMathOperator{\clipping}{clip}
\newcommand{\quant}{\quantize}
\newcommand{\clip}[3]{\clipping_{#1}^{#2}\left(#3\right)}
\newcommand{\round}{\text{round}}

\begin{document}

\begin{acronym}
\acrodef{AIMC}[AIMC]{Analog In-Memory Computing}
\acrodef{ADC}[ADC]{analog-to-digital converter}
\acrodef{DAC}[DAC]{digital-to-analog converter}
\acrodef{DPU}[DPU]{Digital Processing Unit}
\acrodef{AIHWKIT}[AIHWKit]{Analog Hardware Acceleration Kit}
\acrodef{DNN}[DNN]{Deep Neural Network}
\acrodef{PCM}[PCM]{Phase Change Memory}
\acrodef{RRAM}[ReRAM]{Resistive Random Access Memory}
\acrodef{MRAM}[MRAM]{Magnetic Random Access Memory}
\acrodef{ECRAM}[EcRAM]{Electrochemical Random Access Memory}
\acrodef{NVM}[NVM]{Non-Volatile Memory}
\acrodef{IMC}[IMC]{In-Memory Computing}
\acrodef{HWA}[HWA]{hardware-aware}
\acrodef{D2D}[D2D]{Device-to-Device}
\acrodef{C2C}[C2C]{Cycle-to-Cycle}
\acrodef{BL}[BL]{Bit-Line}
\acrodef{WL}[WL]{Word-Line}
\acrodef{RTN}[RTN]{Random Telegraph Noise}
\acrodef{MVM}[MVM]{Matrix-Vector Multiplication}
\acrodef{ML}[ML]{Machine Learning}
\acrodef{LSTM}[LSTM]{Long Short-Term Memory}
\acrodef{SGD}[SGD]{Stochastic Gradient Descent}
\acrodef{MP}[MP]{Mixed-Precision}
\acrodef{TT}[TT]{Tiki-Taka}
\acrodef{TTv2}{Tiki-taka II}
\acrodef{cTTv2}[c-TTv2]{Chopped-TTv2}
\acrodef{AGAD}{Analog Gradient Accumulation with Dynamic reference}
\acrodef{FP}{floating point}
\acrodef{3FC}{3-layered fully-connected}
\acrodef{AAICC}{Analog AI Cloud Composer}

\end{acronym}
\title{Using the IBM Analog In-Memory Hardware Acceleration Kit for Neural Network Training and Inference} 
\author{Manuel Le Gallo}\affiliation{IBM Research Europe, 8803 R\"{u}schlikon, Switzerland}
\author{Corey Lammie}\affiliation{IBM Research Europe, 8803 R\"{u}schlikon, Switzerland}
\author{Julian B\"{u}chel}\affiliation{IBM Research Europe, 8803 R\"{u}schlikon, Switzerland}
\author{Fabio Carta}\affiliation{IBM Research - Yorktown Heights, NY 10598, USA}
\author{Omobayode Fagbohungbe}\affiliation{IBM Research - Yorktown Heights, NY 10598, USA}
\author{Charles Mackin}\affiliation{IBM Research - Almaden, San Jose, CA 95120, USA}
\author{Hsinyu Tsai}\affiliation{IBM Research - Almaden, San Jose, CA 95120, USA}
\author{Vijay Narayanan}\affiliation{IBM Research - Yorktown Heights, NY 10598, USA}
\author{Abu Sebastian}\affiliation{IBM Research Europe, 8803 R\"{u}schlikon, Switzerland}
\author{Kaoutar El Maghraoui}\email{kelmaghr@us.ibm.com}\affiliation{IBM Research - Yorktown Heights, NY 10598, USA}
\author{Malte J. Rasch}\email{malte.rasch@ibm.com}\affiliation{IBM Research - Yorktown Heights, NY 10598, USA}

\date{\today}

\begin{abstract}
\ac{AIMC} is a promising approach to reduce the latency and energy consumption of \ac{DNN} inference and training. 
However, the noisy and non-linear device characteristics, and the non-ideal peripheral circuitry in \ac{AIMC} chips, require adapting \acp{DNN} to be deployed on such hardware to achieve equivalent accuracy to digital computing. In this tutorial, we provide a deep dive into how such adaptations can be achieved and evaluated using the recently released IBM \ac{AIHWKIT}, freely available at \url{https://github.com/IBM/aihwkit}. The \ac{AIHWKIT} is a Python library that simulates inference and training of DNNs using \ac{AIMC}. We present an in-depth description of the \ac{AIHWKIT} design, functionality, and best practices to properly perform inference and training. We also present an overview of the Analog AI Cloud Composer, a platform that provides the benefits of using the \ac{AIHWKIT} simulation in a fully managed cloud setting along with physical \ac{AIMC} hardware access, freely available at \url{https://aihw-composer.draco.res.ibm.com}. Finally, we show examples on how users can expand and customize \ac{AIHWKIT} for their own needs. This tutorial is accompanied by comprehensive Jupyter Notebook code examples that can be run using \ac{AIHWKIT}, which can be downloaded from \url{https://github.com/IBM/aihwkit/tree/master/notebooks/tutorial}.
\end{abstract}


\maketitle 

\section{Introduction}\label{sec:introduction}
\acresetall
Despite providing remarkable breakthroughs in various domains, \acp{DNN} have been accompanied by a dramatic and growing increase in computational demands for training and inference. With the slowing down of Moore’s law and the ending of Dennard scaling, power consumption becomes a key design constraint. Thus, energy-efficient implementations on emerging specialized hardware that leverage approximate and in-memory computing techniques have become essential for AI systems. This has been accompanied by a rise in dedicated AI hardware accelerators, and an increased interest in AI processors that are efficient or fast, or both, when carrying out AI tasks. In addition to traditional digital accelerators, including the Google Tensor Processing Unit, Amazon Inferentia, and IBM Artificial Intelligence Unit~\cite{AIU}, accelerators based on \ac{AIMC} using \ac{NVM} are being actively researched~\cite{Y2020sebastianNatNano,lanza2022,shimengCAS2021}. \ac{AIMC} accelerators that are based on resistive memory device technologies such as \ac{PCM}\cite{Y2022-khaddam-aljameh-JSSC,ares2021,legallo2023,ambrogio2023analog}, \ac{RRAM}\cite{wan2022compute,hung2021four,Zhang2023,Hung2023}, and \ac{MRAM}\cite{deaville202222nm}, have shown great promise in accelerating and reducing the power consumption of deep learning systems. By leveraging the physical properties of such memory devices, computations are performed at the same place where the data is stored, which could considerably improve the run-time and power consumption over today's digital computing technology \cite{jain2023}. 
In an \ac{AIMC} chip, spatially instantiated synaptic weights are encoded in the tunable analog conductance of these devices arranged in crossbar arrays. \acp{MVM} are amongst the most ubiquitous operations in deep learning, and can be performed directly using the network weights stored on the chip \cite{Y2017burrAPX}. Additionally, weight updates for \ac{DNN} training can be performed in-place by tuning the device conductance with suitable programming pulses \cite{gokmen2016pulsed,Agarwal2017}. 

However, despite prolonged ongoing efforts, analog resistive memory devices suffer from various nonidealities, such as device-to-device and cycle-to-cycle variations. These inherent characteristics limit their accuracy and reliability to use in practical deep learning workloads \cite{Y2020NandakumarIEDM,legallo2022,mackin2022optimised}. Therefore, many large-scale simulations encompassing device and circuit nonidealities have been performed to quantify their impact on \ac{DNN} accuracy for training and inference \cite{Joshi2020,rasch2023hardware,liu2022memristor,marinella2020,neurosim_validation,Lammie2022a,nvsim,mnsim}. Although some of these studies have been realized on circuit-level simulators (e.g. SPICE), the size and complexity of deep learning workloads motivated the adoption of an alternative approach of using customized simulation frameworks/toolkits, which are integrated into modern deep learning frameworks, including PyTorch and TensorFlow. In contrast to SPICE-based simulation, which is cycle-accurate, this new alternate approach provides an interface between accurate mathematical models of non-ideal device characteristics and peripheral circuitry, and high-level deep learning frameworks. This methodology enables seamless integration between modern \ac{DNN} frameworks and the noisy physical characteristics of \ac{AIMC} hardware, by modeling the physical properties of \ac{AIMC}, and taking them into account for the training and inference of state-of-the-art \ac{DNN} models. It is within this scope that we have recently open-sourced the IBM \ac{AIHWKIT}, a simulation toolkit that focuses on the algorithmic and functional levels, as opposed to hardware and circuit design levels~\cite{aihwkit}. The aim of this toolkit is to provide a complete software package to estimate the accuracy of \acp{DNN} mapped to \ac{AIMC} hardware, for the advancement of algorithmic analog deep learning.

In \tabref{related_work}, we compare key features of the \ac{AIHWKIT} to related open-source \ac{AIMC} simulation toolkits. Traditional, i.e., SPICE-based simulators, are not compared. We refer the reader to~\cite{Lammie2022a} for a more comprehensive overview. As listed in \tabref{related_work}, only three out of the listed five toolkits are actively maintained: NeuroSim, \ac{AIHWKIT}, and CrossSim. The toolkits are compared against five key dimensions: ML library, supported network types, on-chip inference capabilities, on-chip training, and on-chip inference. Despite its current lack of support for performance estimation, the \ac{AIHWKIT} is the only actively maintained tool which supports all the features listed, and fully embraces modernized software engineering practices. In addition to being available on popular package indexes (PyPi and conda-forge\footnote{CPU and GPU versions can be installed from \url{https://anaconda.org/conda-forge/aihwkit} and \url{https://anaconda.org/conda-forge/aihwkit-gpu}, respectively.}), the \ac{AIHWKIT} uses automated continuous integration and continuous development services (CI/CD) (e.g., Travis) to execute unit tests, and to build and deploy standardized packaged releases.

It is noted that a large number of \ac{AIMC} simulation frameworks have been developed. However, most of them remain closed-source or have been solely used for standalone research projects. Hence, they have not attracted significant attention from the broader research community. Consequently, they have been omitted from our comparative study. While many of these toolkits are complementary in nature, such as those listed in \tabref{related_work}, it is clear that the lack of standardization and excessive tool fragmentation are still prevalent when it comes to \ac{AIMC} simulation and software toolkits.

\begin{table}[!t]
\centering
\caption{Comparison of the \ac{AIHWKIT} with different related open-source \ac{AIMC} simulation frameworks/toolkits. Traditional SPICE-based simulators are not compared.}\label{tab:related_work}
\begin{threeparttable}
  \definecolor{Gray}{gray}{0.95}
  \newcolumntype{g}{>{\columncolor{Gray}}c}

  \begin{tabular}{ll|cccgcc} 
    \toprule
    \multicolumn{2}{l|}{\thead[l]{Framework}}
    
    & \makecell{NeuroSim\\and\\Derivatives \\\cite{Chen2017,Chen2018,Luo2019,Peng2019,Peng2021}}
    & \makecell{XB-SIM~\cite{Liu2019}}
    & MemTorch~\cite{Lammie2020,Lammie2022}
    & \thead{IBM Analog\\ Hardware \\ Acceleration\\ Kit~\cite{aihwkit}}
    & CrossSim~\cite{Xiao2022} \\ \midrule
    
    \multicolumn{2}{l|}{\thead[l]{Year}} & 2017 & 2019 & 2020 & 2021 & 2022 \\ \midrule
    
    \multicolumn{2}{l|}{\thead[l]{Prog. Language(s)}}
    & \makecell{Python, \\C, C++}
    & \makecell{Python, \\C++, \\CUDA}
    & \makecell{Python, \\C, C++,\\CUDA}
    & \Gape[0pt][2pt]{\makecell{Python, \\C++, \\CUDA}}
    & \makecell{Python,\\CuPy} \\
    \midrule

    
    \multirow{2}{*}[-15pt]{\rotcell{\makecell[c]{ML \\ Library}}}
    & \thead[l]{PyTorch} & \checkmark  & & \checkmark & \checkmark &  \\
    & \thead[l]{TensorFlow} & \checkmark &  &  &  & \checkmark \\ \midrule

    \multirow{4}{*}[-40pt]{\rotcell{\makecell[c]{Supported \\ Network Types}}}

    & \thead[l]{Dense (MLP)\tnote{1}} & \checkmark  & \checkmark & \checkmark & \checkmark & \checkmark \\
    & \thead[l]{Convolutional} & \checkmark  & \checkmark & \checkmark & \checkmark & \checkmark \\
    & \thead[l]{Recurrent} &  &  & \checkmark & \checkmark &  \\
    & \thead[l]{Transformer} &  &  & & \checkmark &  \\ \midrule

    \multirow{5}{*}[-16pt]{\rotcell{\makecell[c]{On-Chip \\ Inference}}} 
    & \thead[l]{Accuracy Est.} & \checkmark  & \checkmark & \checkmark & \checkmark & \checkmark \\    
    &\thead[l]{HW-Calib. Noise}& \checkmark  &  &  & \checkmark & \checkmark \\ 
    &\thead[l]{HWA Training\tnote{2}} & \checkmark &  &  & \checkmark & \\ 
    &\thead[l]{Performance Est.} & \checkmark & \checkmark &  &  & \\\midrule
    
    \multirow{4}{*}[-14pt]{\rotcell{\makecell[c]{On-Chip \\ Training}}} 
    & \thead[l]{Digital Gradient} & \checkmark  &  &  & \checkmark & \checkmark \\
    & \thead[l]{In-memory Grad.} &   &  &  & \checkmark &  \\ 
    &\thead[l]{Performance Est.} & \checkmark &  &  &  & \\ 
    \midrule

    \multicolumn{2}{l|}{\thead[l]{Unit Testing}} &  &  & \checkmark & \checkmark &  \\ \midrule
    \multicolumn{2}{l|}{\thead[l]{Package Index(s)}} &  &  & PyPi & PyPi, CF\tnote{3} &  \\ \midrule
    \multicolumn{2}{l|}{\thead[l]{Actively Maintained\tnote{4}}} & \checkmark &  &  & \checkmark & \checkmark \\
    \bottomrule 
\end{tabular}

\begin{tablenotes}
\item[1] Multi-Layer Perceptron.
\item[2] Hardware-Aware Training.
\item[3] Conda-Forge.
\item[4] As per the current date of publication.
\end{tablenotes}
\end{threeparttable}
\end{table}

The rest of the paper is organized as follows. In section~\ref{sec:analogai}, \ac{AIMC} concepts are introduced to familiarize the reader with the kind of research problems that can be tackled with \ac{AIHWKIT}. In section~\ref{sec:aihwkit}, a comprehensive overview of the \ac{AIHWKIT} design is provided, along with a detailed description of each simulated \ac{AIMC} nonideality. Then, in sections~\ref{sec:inference} and~\ref{sec:training}, in-depth step-by-step descriptions on how to perform inference and training with \ac{AIHWKIT} are provided. We explain standard practices to faithfully capture hardware aspects as well as algorithmic techniques to improve accuracy. In section~\ref{sec:composer}, we present the Analog AI Cloud Composer that leverages the \ac{AIHWKIT} simulation platform to allow a seamless no-code interactive cloud-hosted experience and provide physical \ac{AIMC} hardware access. In section~\ref{sec:extension}, we provide three concrete examples of customization of \ac{AIHWKIT} that the user could implement to fit their own research needs. Finally, section~\ref{sec:outlook} provides an outlook on possible future research directions and additions for \ac{AIHWKIT}. 

\clearpage
\section{AIMC Concepts}\label{sec:analogai}
\subsection{Detailed Introduction to AIMC}

By exploiting the physical attributes of memory devices and their array-level organization, it is possible to perform specific computational tasks in the memory itself without the need to shuttle data between the memory and the processing units. The \ac{AIMC} computational paradigm is paving the way for a range of applications, including scientific computing and deep learning\cite{Y2020sebastianNatNano}. Memory devices exhibiting two or more stable states can perform in-memory arithmetic operations such as \acp{MVM}. For example, to perform the matrix-vector multiplication $W\mathbf{x} = \mathbf{y}$, the elements of matrix $W$, i.e. $w_{ij}$, can be mapped linearly to the conductance values of memory-based unit-cells organized in a crossbar configuration. The values of the input vector $\mathbf{x}$ can be mapped linearly to the amplitudes (durations) of read voltages, applied to the crossbar along the rows, or \acp{WL}. The resulting current (charge) measured along the columns of the array, or \acp{BL}, will be proportional to the result of the computation, $\mathbf{y}$. Another attribute exploited for computation is accumulative behavior, whereby the device conductance progressively increases or decreases with the successive application of programming pulses. This enables tuning of the synaptic weights of a neural network during training.

\begin{figure}[t!]
\centering
\begin{tabular}{c}
\includegraphics[width = \columnwidth]{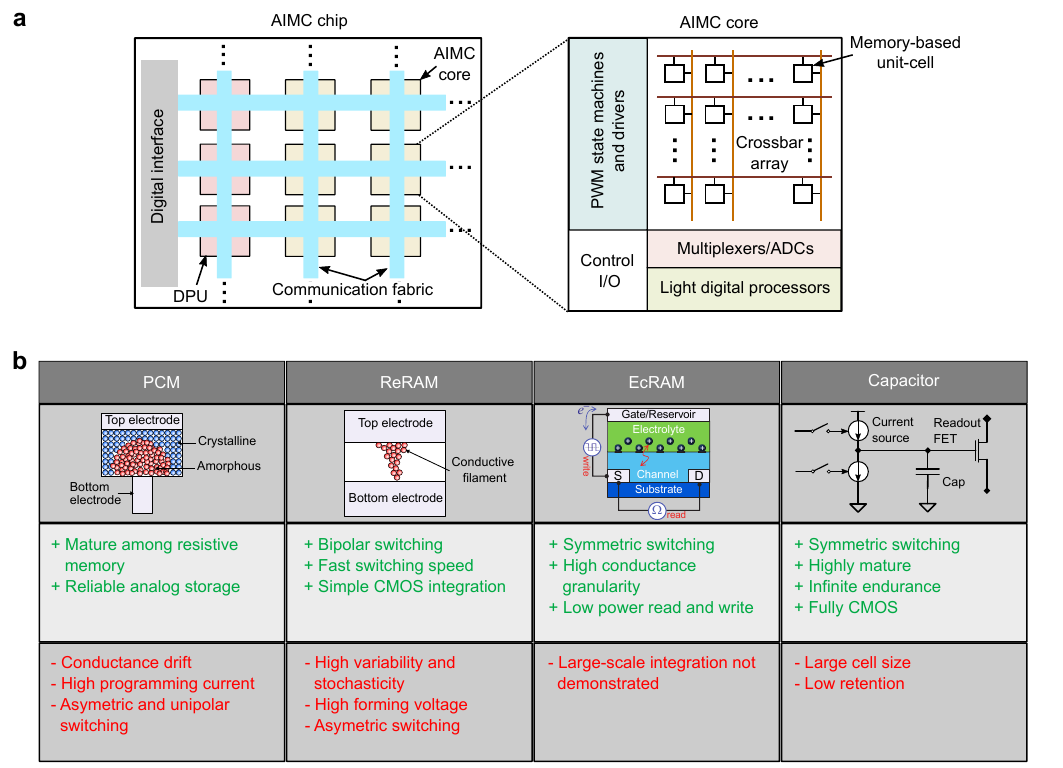}
\end{tabular}
\caption{(a), Illustration of a potential \ac{AIMC} chip. (b) \ac{AIMC} devices implemented in \ac{AIHWKIT} and their properties. } \label{fig:AIMC_chip}
\end{figure}

As shown in Fig.~\ref{fig:AIMC_chip}(a), an \ac{AIMC} chip would ideally comprise a network of \ac{AIMC} cores, each of which would perform a \ac{MVM} primitive along with some light digital post-processing operations. Each \ac{AIMC} core comprises a crossbar array of memory-based unit-cells along with the bit-line drivers, \acfp{ADC}, custom digital compute units to post-process the raw \ac{ADC} outputs, local controllers, transceivers, and receivers. Core-to-core communication can be realized using a flexible on-chip network, akin to those used in traditional digital \ac{DNN} accelerators. To realize a complete \ac{AIMC} accelerator for \ac{DNN} workloads, \ac{AIMC} cores that each perform weight-stationary and energy-efficient \ac{MVM} operations at $\mathcal{O}(1)$ time complexity can be combined with special-function \acp{DPU} to implement auxiliary \ac{DNN} operations, such as activation functions and self-attention compute. Such an architecture is projected to provide highly competitive throughput while offering 40x-140x higher energy efficiency than an NVIDIA A100 GPU \cite{jain2023}. Therefore, there is a strong premise for \ac{AIMC} to enable highly efficient execution of \ac{DNN} workloads. 

There are many promising candidates for the memory element in \ac{AIMC}, including \ac{PCM}, \ac{RRAM}, \ac{ECRAM}, complementary metal-oxide semiconductor (CMOS) capacitive cells, Flash memory, \ac{MRAM}, ferroelectric memory such as ferroelectric field-effect transistor (FeFET) or ferroelectric tunnel junction (FTJ), and photonic memory. The list shown in \figref{AIMC_chip}(b) is not a complete list of possible memory elements, but provides examples of how analog resistance levels are achieved with various materials and circuit implementations. All devices described in \figref{AIMC_chip}(b) have hardware-calibrated models implemented in \ac{AIHWKIT} to simulate training and/or inference (see Sections \ref{sec:inference} and \ref{sec:training}). In \ac{PCM}, data is stored by using the electrical resistance contrast between a high-conductive crystalline phase and a low-conductive amorphous phase of a phase-change material. The phase-change material resistance can be modulated by creating amorphous regions of varying sizes through the application of electrical current pulses. \ac{RRAM} switches between high and low conductance states based on the formation and dissolution of a filament in a non-conductive dielectric material. Intermediate conductance is achieved either by modulating the width of the filament or by modulating the composition of the conductive path. \ac{ECRAM} modulates the conductance between source and drain terminals using the gate reservoir voltage that drives ions into the channel. Lastly, CMOS-based capacitive cells can also be used as memory elements for analog computing, as long as leakage is controlled and performing the compute and read operations can be completed quickly. 

Clearly, at the time of writing, there is still no ``optimal'' \ac{AIMC} device technology, as each one of the current available technologies has its strengths and weaknesses, as illustrated in \figref{AIMC_chip}(b). For instance, \ac{PCM} devices are arguably considered the most mature among resistive memory types, however they suffer from temporal conductance drift, and the uni-polar/asymmetric switching behaviour leads to several complications for training. This is one of the key motivations behind building a simulator like the \ac{AIHWKIT}, as to allow the exploration of the impact of various devices with their multitude of characteristics on the performance of AI models.

\subsection{How to Perform DNN Training and Inference with AIMC}

\begin{figure}[t!]
\centering
\begin{tabular}{c}
\includegraphics[width = \columnwidth]{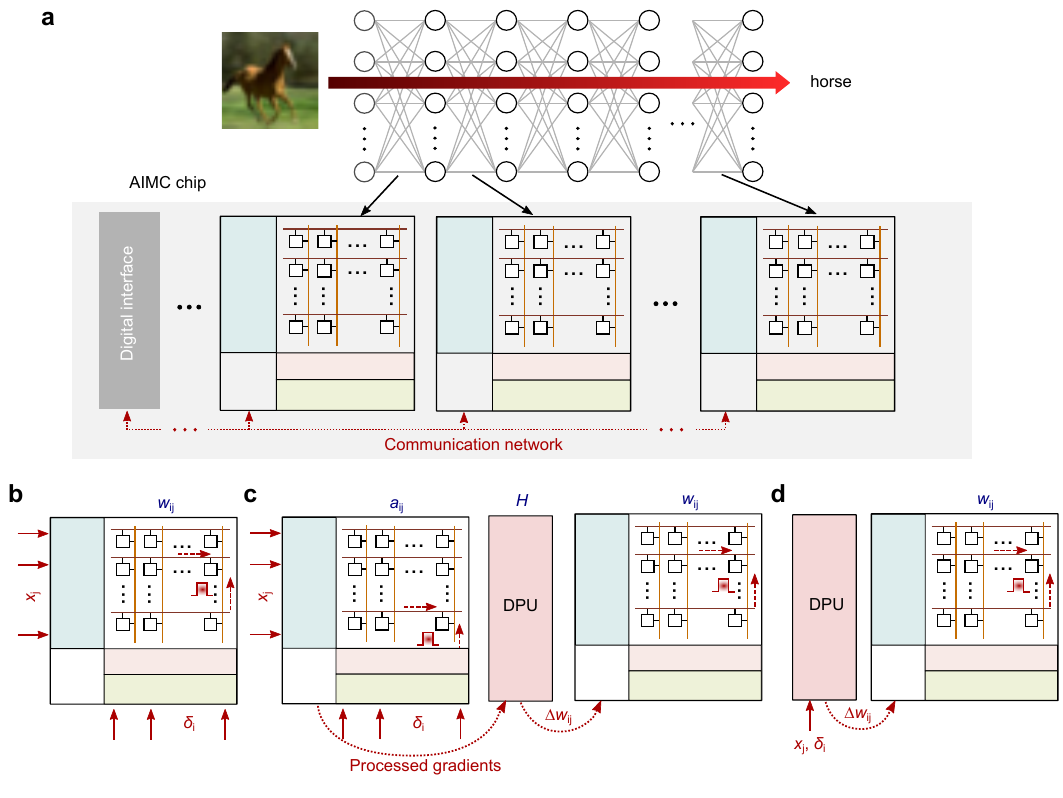}
\end{tabular}
\caption{(a) Mapping of a neural network to an \ac{AIMC} chip. (b) Implementation of in-memory SGD weight update. (c) Implementation of \ac{TTv2} weight update. (d) Implementation of mixed-precision weight update. } \label{fig:inference_training}
\end{figure}

A neural network layer can be implemented on (at least) one crossbar array of an \ac{AIMC} core, in which the weights of that layer are stored in the charge or conductance state of the memory devices at the crosspoints (see \figref{inference_training}(a)). Because the state of a memory device can encode only a positive quantity, usually at least two devices in a differential configuration are used per weight: one to represent a positive synaptic weight component and the other one to represent a negative weight component. The propagation of data through the layer is performed in a single step by inputting the data to the crossbar rows and deciphering the results at the columns. The results are then passed through the neuronal activation function and input to the next layer. The neuronal activation function is typically implemented at the crossbar periphery, using analog or digital circuits. Because every layer of the network is stored physically on different arrays, each array needs to communicate at least with the array(s) storing the next layer for feed-forward networks, such as multi-layer perceptrons (MLPs) or convolutional neural networks (CNNs). For recurrent neural networks (RNNs), the output of an array needs to communicate with its input.

The efficient matrix multiplication realized via \ac{AIMC} is very attractive for inference-only applications, where data is propagated through the network on offline pre-trained weights. In this scenario, the weights are typically trained using conventional GPU-based hardware, and then are subsequently programmed into the \ac{AIMC} chip which performs inference. However, because of device and circuit level nonidealities in the \ac{AIMC} chip, custom techniques must be included into the training algorithm to mitigate their effect on the network accuracy (so-called \ac{HWA} training). For inference tasks, device nonidealities that affect the network accuracy include conductance drift, programming errors, read noise and stuck on/off devices. Circuit nonidealities, including finite resolution of \acp{DAC} and \acp{ADC}, parasitic voltage drops on the devices during readout when a high current is flowing through the crossbar wires (IR-drop), noise from the peripheral circuits at the crossbar output (e.g. amplifiers), and parasitic currents from sneak-paths during readout will also negatively impact the accuracy.

\ac{AIMC} can also be used in the context of neural network training with backpropagation. This training involves three stages: forward propagation of labelled data through the network, backward propagation of the error gradients from output to the input of the network, and weight update based on the computed gradients with respect to the weights of each layer. This procedure is repeated over a large dataset of labelled examples for multiple epochs until satisfactory performance is reached by the network. When performing training of a neural network mapped on \ac{AIMC} cores, forward propagation is performed the same way as inference as described above. The only difference is that all the activations $x_j$ of each layer have to be stored locally in the periphery. Next, backward propagation is performed by inputting the error gradient $\delta_i$ from the subsequent layer onto the columns of the current layer and deciphering the result from the rows. The resulting sum $\sum_i \delta_iw_{ij}$ needs to be multiplied by the derivative of the neuron nonlinear function, which is computed externally, to obtain the error gradient of the current layer. Finally, weight updates are implemented based on the outer product of activations and error gradients $\Delta w_{ij} \propto \delta_ix_j$ of each layer. 

The weight update is performed in-memory by applying suitable electrical pulses to the devices which will increase or decrease their conductance in proportion to the desired weight update. There are multiple approaches to perform the weight update with \ac{AIMC}. Each approach has its advantages and drawbacks. One approach is to perform a parallel weight update by sending deterministic or stochastic overlapping pulses from the rows and columns simultaneously to implement an approximate outer product and program the devices at the same time (\figref{inference_training}(b))~\cite{gokmen2016pulsed,Agarwal2017,Y2015burrITED}. This method, which we term in-memory stochastic gradient descent (in-memory SGD), has the advantage to perform a fully-parallel analog weight update on the crossbar array at O(1) time complexity, and therefore is highly efficient in terms of speed. However, it requires stringent specifications on the conductance update granularity (minimum increase/decrease of device conductance with a single pulse), asymmetry (difference in device response when increasing or decreasing conductance) and linearity (dependence of conductance update on the device conductance state) to obtain accurate training, and high device endurance is critical. To mitigate some of these issues, the Tiki-Taka training algorithm was proposed~\cite{gokmen2020,gokmen2021}, which significantly relaxes the device conductance update requirements. Here, two matrices are encoded in \ac{AIMC} cores, $A$ and $W$. $W$ encodes the network weights, whereas $A$ computes and accumulates the weight gradient information. $A$ is updated via parallel weight updates as described for in-memory SGD. After a certain number of updates on $A$, $W$ is updated based on reading the gradient information from $A$ via parallel weight updates. In the second version of Tiki-Taka (\ac{TTv2}),~\cite{gokmen2021,rasch2023fast} an additional matrix $H$, implemented in the digital domain, is used. $H$ implements a low pass filter while transferring the gradient information processed by $A$ to $W$, which further improves the robustness to nonideal conductance updates. This low pass filter reduces the gradient noise and averages the gradient information over more inputs before updating the weights. A schematic implementation of the \ac{TTv2} weight update is shown in \figref{inference_training}(c). Finally, a third approach is to perform so-called mixed-precision deep learning, by computing the weight updates in a separate digital processor and accumulating them in a high-precision digital memory (\figref{inference_training}(d))~\cite{nandakumar2020}. When the accumulated weight updates reach a threshold, the corresponding devices get updated through single-shot programming pulses. This approach is much less sensitive to nonidealities such as limited device granularity because the gradient is not computed using \ac{AIMC} but instead in high-precision \ac{FP}.  It is also more flexible since more complex learning rules can readily be implemented in digital. Moreover, the digital computation and accumulation of weight updates significantly relax the requirements on device endurance. However, the cost of the digital computations is significant (${\cal O}(n^2)$ for a $n \times n$ weight matrix), and thus limits the speed of the \ac{AIMC} training, even though forward and backward passes are fast (${\cal O}(1)$).  In contrast, for the in-memory SGD and Tiki-taka learning rules, the number of additional digital operations is linear to the size of the input vector (${\cal O}(n)$) and often executed only periodically, so that the update is done much faster than for mixed-precision. All three methods presented here, as well continuously improved versions,  are implemented in \ac{AIHWKIT}, and section~\ref{sec:training} describes how to configure them for testing on different \ac{AIMC} device models. 

\clearpage
\section{AIHWKIT design}\label{sec:aihwkit}
As laid out in the previous section, \ac{AIMC} can accelerate certain parts of typical DNN (and other computing) workloads. Dense \acp{MVM} are particularly favorable for \ac{AIMC}, when the matrix elements are stationary and stored in (analog) memory. However, today's \acp{DNN} are often heterogeneous and include a variety layers, such as non-linear activation functions or attention mechanisms, that cannot be efficiently computed in-memory. The \ac{AIHWKIT}, which primarily focuses on functional verification of \ac{AIMC}, is thus designed to handle both digital as well as \ac{AIMC} components within the same \ac{DNN} compute graph.               

\subsection{Simulator Code-design Overview}
\begin{figure}
    \centering
    \includegraphics[width=0.7\textwidth]{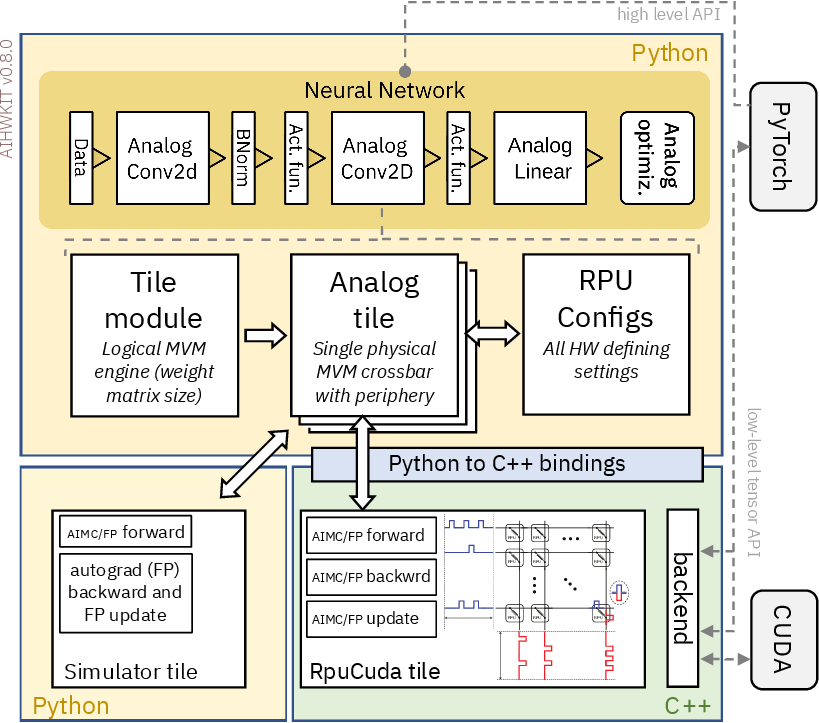}
    \caption{Design of the \ac{AIHWKIT}. A DNN is defined with typical \pytorch\ commands, except for layers that are to be performed in \ac{AIMC}. We provide analog layers to implement convolution layers, linear layers etc. (see \tabref{analog-layers}). Each of these analog layer modules contain (at least) one analog tile module that encapsulates the analog computations as well as concatenating of logical tile arrays. Each analog tile module consists of one or multiple \emph{analog tiles}. These analog tiles encapsulate the \ac{NVM} crossbar operations together with immediate peripheral compute (such as \ac{ADC} and \ac{DAC}, affine output scaling and bias). Each analog tile can be configured in a broad way using a \RPUConfig. The \RPUConfig\ determines in a highly customizable way how the nonideal \ac{AIMC} forward, backward, and update behaviour is actually implemented and what peripheral aspects and device materials are used in the \ac{AIMC} hardware of investigation. }
    \label{fig:design}
\end{figure}

Since the \ac{AIHWKIT} is based on the ML framework \pytorch, the user can rely on the vast library of digital \ac{FP} layers and functions for defining common \acp{DNN}. Only some layers of the \ac{DNN} that are supposed to run on \ac{AIMC} will use the simulation \ac{AIMC} capabilities of the \ac{AIHWKIT}. The overall design is depicted in \figref{design}. The \ac{DNN} is defined conveniently in standard \pytorch\ syntax using e.g. using \longvar{Linear} and \longvar{Conv2d} layer modules for fully-connected and convolution layers, respectively. If one decides to simulate such a layer on \ac{AIMC}, \ac{AIHWKIT} provides corresponding layer modules, such as \longvar{AnalogLinear} and \longvar{AnalogConv2d}, respectively, that simulate the underlying matrix-vector products with customizable \ac{AIMC} nonidealities. In such a way, the impact of \ac{AIMC} nonidealities on the function of the DNN (e.g. prediction accuracy) can be measured. The analog layers available are listed in \tabref{analog-layers}. As illustrated further in \figref{design}, each analog layer module consists of one or multiple \emph{analog tiles}, that are meant to be a single physical crossbar core with immediate periphery. Analog weights are assumed stationary once initialized. For instance, a large linear layer could be made up of multiple $512 \times 512$ crossbar arrays, where multiple non-ideal \acp{MVM} need to be performed and concatenated. The partial sum of the individual outputs is assumed to be computed in \ac{FP} accuracy. In this case, the \emph{tile module} would consist of multiple \emph{analog tiles} with additional digital summations. Each analog tile in \ac{AIHWKIT} itself consists of a physical (simulated) memristive crossbar (of class \longvar{SimulatorTile}), as well as immediate periphery such as \acp{ADC}, or error dynamic corrective steps, such as noise or bound management~\cite{Gokmen2017}. Depending on the hardware customization it can also hold an affine transform (digital output scales and biases, global or column-wise), which is known to greatly improve the mapping of weights to conductances, and is needed for converting ADC-tics to meaningful quantities for the subsequent layers of the DNN~(see also~\cite{rasch2023hardware}).

In \ac{AIHWKIT}, nonidealities, material response characterization, and general hardware configurations of each analog tile can be specified by a \RPUConfig. The \RPUConfig is in principle unique per analog tile, however, in common use cases one assumes the same \RPUConfig\ for each analog tile on the chip. We will explain how to configure the \ac{AIMC} hardware using the \RPUConfig\ in detail in the next section. Internally, each analog tile will call the low-level \texttt{SimulatorTile} class for actually performing the non-ideal computations requested by the \RPUConfig. As indicated in \figref{design}, a number of optimized core routines are available that simulate the \ac{AIMC} \acp{MVM}. In particular for analog training, when the \ac{MVM} as well as the outer-product update are both done in-memory, the C++/CUDA library ({\it RPUCuda}) is used through python bindings to increase the simulation performance.

\begin{table}[!t]
    \centering
    \caption{Analog layer modules. Additionally, the toolkit provides \texttt{Mapped} versions that enforce the mapping of large weight matrices onto multiple physical tiles.}\label{tab:analog-layers}
    \begin{tabular}{lrr}
        \toprule \toprule
         \textbf{Module Name} & \textbf{Torch Equivalent} & \textbf{Functionality}\\
         \midrule 
         \texttt{AnalogLinear} & \texttt{Linear} & Linear layer with bias\\
        \texttt{AnalogConv1d} & \texttt{Conv1d} & 1-dim convolution\\   
        \texttt{AnalogConv2d} & \texttt{Conv2d} & 2-dim convolution\\   
         \texttt{AnalogConv3d} & \texttt{Conv3d} & 3-dim convolution\\
        \texttt{AnalogRNN} & \texttt{RNN} & Recurrent layer(s) with configurable cell \\    
         \texttt{AnalogLSTM} & \texttt{LSTM} & Uni/bi-directional LSTM layers \\  
         \bottomrule \bottomrule
    \end{tabular}
\end{table}

\subsection{Model Conversion and Analog Optimizers}
As described in the previous section, typical \pytorch\ syntax is used to define the DNN to be simulated. This has the advantage that the vast amount of pre-coded and DNNs available for download from the ML community are readily usable for \ac{AIHWKIT}. However, layers within the \acp{DNN} that should run using \ac{AIMC} need to be replaced by their ``analog'' counterpart (see \tabref{analog-layers}). To ease the conversion of pre-coded (and possibly pre-trained) DNNs to \ac{AIHWKIT}, convenient conversion tools are provided that automatically replace \pytorch\ layers, such as \longvar{Linear},  with their counterparts, e.g. \longvar{AnalogLinear}. Thus e.g. a call 
\begin{minted}{python}
    from aihwkit.nn.conversions import convert_to_analog
    analog_model = convert_to_analog(model, rpu_config) 
\end{minted}
would convert all applicable layers of the \ac{FP} DNN to an \emph{analog model} featuring \ac{AIMC} layers, where all analog tiles instantiated are configured using the same hardware configuration \RPUConfig\ and the \ac{FP} weights. Note that here we always assume that enough analog tile resources are available on the chip to store the requested weight matrices of the \ac{DNN}. Furthermore, weights are initialized perfectly without any programming noise, which is appropriate for untrained DNNs, as the intial setting is random anyway. However, if weights are pretrained, extra steps are necessary to actually program the weights into the conductances of the crossbars, so that they show realistic deviation from the targets as expected for the material choice. We will describe this process in detail in section~\ref{sec:inference}, where we also describe how inference is performed on this \emph{analog model} and how one would potentially re-train the model with noise injection for increased \ac{AIMC} robustness. 

The \ac{AIHWKIT} provides \emph{analog optimizers}, such as \texttt{AnalogSGD}, that make \pytorch\ aware of the analog layers, so that the correct (custom) forward, backward, and update pass (and potential post-update steps) are performed, as requested in the \RPUConfig. Before going into detail on training and inference, we first introduce the extensive hardware customization possibilities using the \RPUConfig\ in the next sections.    

\begin{table}[t]
    \centering
    \begin{tabular}{lcccp{6cm}}
    \toprule \toprule
         \texttt{RPUConfig} name & Algorithm & Forward & Backward & Update\\
         \midrule 
         {\it Inference}& AIMC inference / SGD & AIMC & FP & FP \\    
        {\it TorchInference} & AIMC inference / SGD & AIMC & FP & FP using \pytorch\ \cite{pytorch} {\it autograd}\\   
          {\it Single}  &  In-memory SGD \cite{gokmen2016pulsed} & AIMC & AIMC & Stoch. pulsed in-memory update ($\pulsed \analog{W}$)\\
         {\it UnitCell} & Specialized SGD & AIMC & AIMC & Using multiple devices (crossbars), based on the compound, see Tab.~\ref{tab:compounds} \\ \bottomrule \bottomrule
          \end{tabular}
    \caption{Examples of different \emph{config} classes (the suffix \texttt{RPUConfig} is omitted).  Note that we make a distinction between chips that are only designed for inference (defined by {\it configs} having \texttt{Inference} in its name) and chips that support in-memory training (all other \texttt{RPUConfig} types). In case of inference-only chips, only the forward pass is done with analog nonidealities and tools are available to add phenomenological programming noise and drift during the evaluation (see \secref{inference}). Training with such a configuration means ``hardware-aware training'', where a more robust \ac{FP} model is trained e.g. with noise injection to be programmed on the analog inference chip during the evaluation phase.  On the other hand, in case of in-memory training,  the backward pass is non-ideal as well and the weight update is defined by the material properties of the device model as pulses will be used to incrementally update the device in-memory using the corresponding gradients. Thus, in this case fully analog in-memory training is performed (see \secref{training} for more details)}
    \label{tab:rpu-configs-types}
\end{table}

\subsection{Tile-level \RPUConfig\ Specifies All Analog Hardware settings} \label{sec:rpu-config}
The \RPUConfig\ is a python data class that has a number of fields and sub-structures which allow the specification of hardware properties, such as the amount and type of nonidealities, in the \ac{AIMC} \acp{MVM}.  On a higher level, \ac{AIHWKIT} provides a number of basic \RPUConfig\ classes that are used to distinguish fundamentally different hardware designs. In particular, it distinguishes between in-memory analog training and chips that do not support training capabilities and instead are used for inference only. Inference-only configurations are based on the \longvar{InferenceRPUConfig} class, whereas in-memory training settings are either derived from the \longvar{SingleRPUConfig} or \longvar{UnitCellRPUConfig} classes (see \tabref{rpu-configs-types} for an overview of different \RPUConfig\ types). Note that the main difference between in-memory training and inference-only chips is how the backward and update nonidealities are defined. While for inference-only chips, they are simply done in \ac{FP} (possibly implementing hardware-aware training, see \secref{inference}), whereas in case of in-memory training configurations allow a plethora of device-material settings and parameters defining specialized \ac{AIMC} \ac{SGD} algorithms.

Given that the \RPUConfig\ mainly specifies the hardware settings, in general all its properties are assumed to be constant and non-changeable after the analog model was constructed using a particular \RPUConfig. However, in some cases, one wants to experiment with one hardware setting e.g. during training, while changing some hardware settings during inference, which would mean to change some properties of the \RPUConfig\ after model creation. While this cannot be done by directly modifying the \RPUConfig\ fields of the constructed model, it still can be done indirectly by exporting and importing of its state as long as the class of the \RPUConfig\ does not change. In more detail, it can be achieved by constructing a second model \longvar{analog_model_new} using a new and modified \RPUConfig\ \longvar{rpu_config_new} and loading the state dictionary from the first model \longvar{analog_model} without loading the \RPUConfig\ from the state dictionary by using the \longvar{load_rpu_config} flag. For example:

\begin{minted}{python}
    analog_model = convert_to_analog(model, rpu_config)
    # [..] e.g. train analog_model here. Then construct new model: 
    analog_model_new = convert_to_analog(model, rpu_config_new)
    analog_model_new.load_state_dict(
        analog_model.state_dict(), load_rpu_config=False)
\end{minted}
Now the new model \longvar{analog_model_new} has the same parameters as \longvar{analog_model} but with a modified \RPUConfig. Any further evaluation or training will thus be based on the new hardware configuration.   

In \tabref{rpu-configs-fields}, typical sub-fields of a \RPUConfig\ are listed. Note that there are other fields that define additional input processing (\longvar{pre_post}) or the weight-to-tile mapping (\longvar{mapping}) properties.  All nonidealities of the \ac{AIMC} \ac{MVM} itself are defined in the \longvar{forward} and \longvar{backward} fields, respectively, as described in the next section.

\begin{table}[t]
    \centering
    \begin{tabular}{llp{8cm}}
    \toprule \toprule
     \RPUConfig\ field & Parameter class & Functionality\\
     \midrule 
      \texttt{tile\_class} & - & Specifies the class used for the analog tile (e.g. \texttt{AnalogTile})\\
      \texttt{tile\_array\_class} & - & Logical array class used if requested (typical \texttt{TileModuleArray})\\
\texttt{device} & \texttt{PulsedDevice} / \texttt{UnitCell} & Specifies the material device properties for in-memory update (e.g. ReRAM-like device-to-device variation during pulsed update)\\
\texttt{forward} & \texttt{IOParameters} & Specify the AIMC MVM nonidealities during the forward pass (e.g. IR drop strength)\\
\texttt{backward} & \texttt{IOParameters} & Specify the AIMC MVM nonidealities during the backward pass (transposed MVM)\\
\texttt{update} & \texttt{UpdateParameters} & Specify the pulsing properties during update (e.g. pulse train length)\\
\texttt{mapping} & \texttt{MappingParameter} & Architectural and peripheral setting (e.g. maximal tile size, whether to use digital affine scales and biases)\\
\texttt{pre\_post} & \texttt{PrePostProcessingParameter} & Pre-post processing (e.g. input range learning)\\
\bottomrule \bottomrule
          \end{tabular}
    \caption{Typical fields of the \RPUConfig\ data class and their functionality. Note that not all fields are available for each of the \RPUConfig\ types (see \tabref{rpu-configs-types}). There are more fields available not mentioned here that are specific to \longvar{InferenceRPUConfig} (such as \longvar{noise_model}, \longvar{drift_compensation}), which specify hardware-aware training and evaluation options for inference-only chip designs (see Sec.~\ref{sec:inference} for a detailed description).   }
    \label{tab:rpu-configs-fields}
\end{table}

\subsection{Configurable MVM Nonidealities}

\begin{table}[!t]
    \centering
    \caption{The \longvar{IOParameters} class customize the \ac{MVM} \ac{AIMC} nonidealities. Here a selection of commonly used settings are summarized. Note that the nonidealities can be selected independently for a "normal" \ac{MVM} during the forward pass and the transposed \ac{MVM}, which is used during the backward propagation in case of in-memory training (see \RPUConfig\ field in \tabref{rpu-configs-fields}).  }
    \begin{tabular}{lcp{11cm}}
    \toprule \toprule 
     Class Field & Typical Value & Functionality\\
     \midrule 
     \longvar{is_perfect} &  False & Debug switch for removing all nonideality settings. \\

\longvar{mv_type} & \textit{OnePass} & Selects the type of analog mat-vec computation. For instance, whether only one pass is performed, so that negative and positive currents are added in analog, or multiple passes, where positive and negative inputs are given sequentially in two passes. \\
     \longvar{noise_management} & \textit{AbsMax} & Type of noise management~\cite{gokmen2016pulsed}, which is a dynamic input scaling per input vector (dynamic quantization).\\

    \longvar{bound_management} & \textit{None} & Type of output bound management. When set to \textit{Iterative}, each \ac{MVM} is "speculatively" computed, which mean that it is dynamically recomputed with reduced inputs only if the output is hit. Note that this incurs a run time penalty in practice.  \\
    \longvar{inp_bound} & 1.0 & Input bound and ranges for the digital-to-analog converter (DAC). The MVM computation is typically normalized to a fixed -1 to 1 input range.\\
    \longvar{ir_drop} &  1.0 &   Scale of IR drop along the inputs (rows of the weight matrix). \\
    
\longvar{w_noise} &  0.01 & Scale of output referred MVM-to-MVM weight read noise.\\
\longvar{w_noise_type}& \textit{AddConstant}& Type of the weight noise for instance additive constant Gaussian to each weight element.\\  
    
\longvar{inp_noise} &  0.0 &  Standard deviation of Gaussian (additive) input noise (after applying the \ac{DAC} quantization)\\
\longvar{inp_res} &  254 & Resolution (or quantization steps) for the full input (signed) range of the \ac{DAC}. \\
\longvar{inp_sto_round} &  False &   Whether to enable stochastic rounding of \ac{DAC}.\\

\longvar{out_bound} & 10.0 &   Output range for analog-to-digital converter (ADC) in normalized units. Typically maximal weight and input is normalized to 1, so that 10 means outputs are clipped at a current generated from 10 max inputs with all max weights. \\  
\longvar{out_noise} &  0.04 &  Standard deviation of Gaussian output noise \\

\longvar{out_nonlinearity} & 0.0 &  S-shaped non-linearity applied to the analog output (with possible output-to-output variation). \\


\longvar{out_res} & 254 &  Number of discretization steps for \ac{ADC} or resolution in the full (signed) output range. \\

\longvar{out_sto_round} & False &   Whether to enable stochastic rounding of \ac{ADC}. \\ \bottomrule \bottomrule
\end{tabular}
    \label{tab:mvm-nonidealities}
\end{table}

As mentioned above, \acp{MVM} implemented on \acp{AIMC} are non-ideal. This is due to a number of device and circuit nonidealities, including, but not limited to: device-to-device and cycle-to-cycle conductance variations, output noise, weight read noise, IR drop, and quantization noise. The \longvar{forward} field of \RPUConfig\ handles attributes related to how each \ac{AIMC} \ac{MVM} is to be performed in the forward pass (during inference as well as during training), and the \longvar{backward} field handles all attributes related to a possibly non-ideal backward pass during  backpropagation. It is noted that all \texttt{forward} or \texttt{backward} attributes \emph{do not change} the underlying weights (conductances) from one \ac{MVM} to the next. Instead, \emph{reversible} noise is added as requested, and for some nonidealities, such as IR-drop, the expected \ac{MVM} output is modified in-place.

Long-term effects, such as diffusion processes, are not considered by default on the level of the duration of processing a single mini-batch. Instead diffusion or decay processes can be applied only after processing a mini-batch. The user has the responsibility to ensure that this approximation of the long-term effects is reasonable for the hardware and materials under investigation. Other long-term weight-related effects, including programming noise, retention, 1/f noise, and drift, can be specified using specialized \RPUConfig\ fields related to inference (e.g. \longvar{noise_model}, see \secref{inference} for details). 

Mathematically, the generally simulated \ac{AIMC} forward and backward passes can be expressed as
\begin{equation}
  \label{eq:mat-vec-analog}
    y_i = \alpha^\text{out}_i f_\text{adc}\big(\sum_j(\analog{w}_{ij} +
      \sigma_\text{w}\xi_{ij})(f_\text{dac}(x_j) 
      + \sigma_\text{inp}\xi_j)  + \sigma_\text{out}\xi_i\big) + \beta_i,
\end{equation}
where $f_\text{adc}$ and $f_\text{dac}$ model the (possible non-linear) analog-to-digital and digital-to-analog processes 
(together with dynamic scaling and range clipping),
and the $\xi$ are Gaussian noise. 
In general, it is assumed to have an analog part of the weight $\analog{W}$, the \emph{analog weight},  that is stored in physical units. The \ac{ADC} counts (that have arbitrary units) are then converted back to the correct \ac{FP} range by a digital out-scaling factor(s) $\alpha^\text{out}_i$ that could either be set to be column-wise (i.e. depending on $i$) or tile-wise. The bias $\beta_i$ could be digital or analog as well. Mathematically, because of the output scales, the actual weight $W$ is given by a combination of the analog weight $\analog{W}$ and the output scales, $W \approx \alpha^\text{out}\analog{W}$. Since the physical units of $\analog{W}$ and $y_i$ are therefore arbitrary (they can be incorporated in $\alpha^\text{out}$), we define the analog weight as well as the input voltage in normalized units (maximally 1) for simplicity, and define all \ac{MVM} nonideality parameters with respect to these normalized units.   

\begin{figure}[!t]
    \centering
    \includegraphics[width=0.5\textwidth]{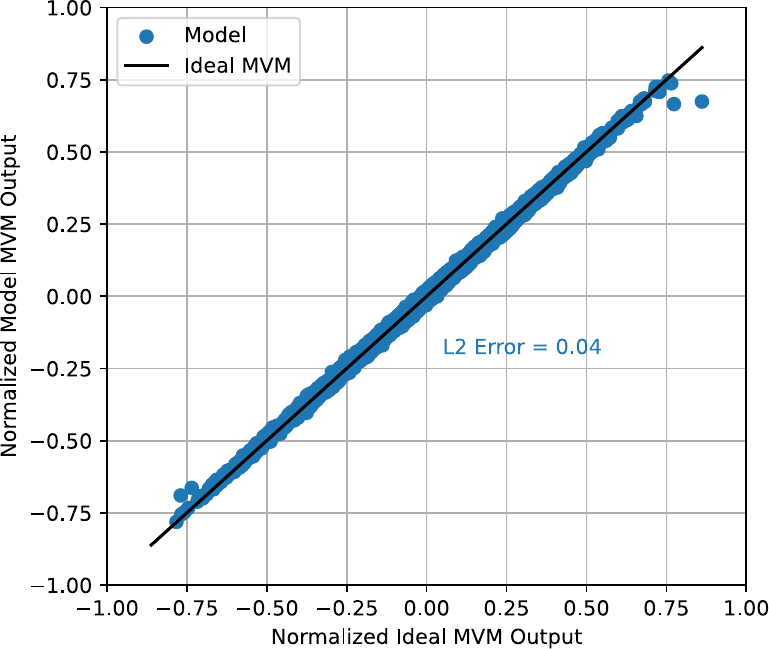}
    \caption{Non-ideal \acp{MVM} from a $512\times512$ analog tile simulated using the \ac{AIHWKIT} with commonly used settings, as listed in Table~\ref{tab:mvm-nonidealities}, when programming noise is not applied. Inputs are sampled from a sparse uniform distribution, with a sparsity of 50\%, and weights are sampled from a clipped Gaussian distribution with a standard deviation of 0.246. Output values are normalized using \longvar{out_bound}, so clipping happens at different normalized output values.}
    \label{fig:typical_non-ideal_MVM}
\end{figure}

A non-ideal \ac{MVM} performed with \ac{AIHWKIT} using the typical value settings shown in \tabref{mvm-nonidealities} is depicted in \figref{typical_non-ideal_MVM}. 
In general, \RPUConfig\ fields can be specified by either passing the keyword values to the \longvar{IOParameter} class, or by simply modifying the attributes of the class. For instance, to set the \longvar{out_noise} parameter of the forward and backward passes, one can write:  
\begin{minted}{python}
    from aihwkit.simulator.configs import SingleRPUConfig, IOParameters
    # choice 1
    rpu_config_1 = SingleRPUConfig(
                       forward=IOParameters(out_noise=0.1), 
                       backward=IOParameters(out_noise=0.1)
                   )
    # choice 2
    rpu_config_2 = SingleRPUConfig()
    rpu_config_2.forward.out_noise = 0.1
    rpu_config_2.backward.out_noise = 0.1
\end{minted}
Note, that while for inference-only chips the forward pass matters (see \secref{inference}), for in-memory training both forward pass and backward \ac{MVM} nonidealities are set separately. Most \RPUConfig\ related classes and enumerators can be imported from \longvar{aihwkit.simulator.configs}. Consequently, we will omit the import statements below. In the following subsections, we give an overview of the different configurable \ac{MVM} nonidealities which can be simulated using the \ac{AIHWKIT}.

\subsubsection{AIMC Network Weight Encoding} \label{sec:mapping}

When performing \acp{MVM}, the conductance of \ac{NVM} elements are usually linearly mapped to a range of weight values, and it is assumed that a typical pulse-width modulation of the voltage input~\cite{Y2022-khaddam-aljameh-JSSC,ares2021} can be approximated by a time average (so that $x$ corresponds to the mean voltage given). Multiple-passes per \ac{MVM} (for example applying positive and negative inputs in two separate phases) can be simulated. However, the toolkit currently does not natively support a bit-wise "digital" mapping of weights, where only 1 and 0 states are (approximately) represented by conductances, and multiple devices are used with different significances to approximate a digital \ac{MVM}~\cite{legallo2022}. However, it could be readily implemented by defining a new analog tile module that consists of multiple analog tiles representing different significances and summing over the individual outputs. In section~\ref{sec:extension} we give some examples of how to customize analog tile modules. 

Before going in more detail describing the simulated \ac{AIMC} nonidealities, there are a number of configurations that define how to map the \ac{FP} weights $W$ to the analog weights $\analog{W}$ and the output scales $\alpha^\text{out}$ (see \eqref{mat-vec-analog}). These are governed by the \longvar{MappingParameter} in the \longvar{mapping} field of the \RPUConfig. 

\paragraph{Analog Tile Size and Bias}
The \longvar{max_in_size} and \longvar{max_out_size} properties set the (maximal) tile size in the input and output dimensions.  If for a given layer the weight matrix is larger than this maximal size, multiple analog tiles will be used to represent the full weight matrix, where the outputs of each tile are assumed to be added up in \ac{FP} precision (after \ac{ADC} conversion) and concatenated and split as demanded. Note that currently there is no accuracy effect of limiting the output size as simulations are all independent for columns. Thus to increase simulation speed it is advisable in most cases to set \longvar{max_out_size} to 0 to turn off the splitting. However, the input size is crucial for some nonidealities (such as IR drops or \ac{ADC} saturation), and thus should be set as required by the hardware design.     

The bias of the analog layer can either be encoded in the analog tile (as an additional column) or assumed to be digital (selected with \longvar{digital_bias}). 

\paragraph{Initial weight mapping}
The property \longvar{weight_scaling_omega} specifies how initially (when (re)setting the weights of an analog layer or using \longvar{analog_tile.set_weights()}) weights $W$ are distributed among the analog weights $\analog{W}$ and the output scale(s) $\alpha^\text{out}$. The value specifies the analog weight value $\analog{w}^*$ that is used for the absolute max of $w_\text{max} \equiv \max_{ij} |w_{ij}|$. Thus for \longvar{weight_scaling_omega} equals $\omega$ (and \longvar{weight_scaling_columnwise=False}), then $\alpha^\text{out}\leftarrow w_\text{max} / \omega$ and $\analog{W}  \leftarrow \omega W / w_\text{max}$. Typically, $\omega=1$ for inference or somewhat smaller for training (see \cite{raschDT2020} for details). This initial weight mapping can also be done per column (thus computing the max per column and having individual output scales per column), when setting \longvar{weight_scaling_columnwise}. 

Note that for the special case $\omega=0$, the initial weight mapping is turned off, that is $\alpha^\text{out} = 1$. In this case, the user has to make sure that \ac{MVM} nonideality values are correctly specified and weights are not too large to invalidate range assumptions. It is advisable to always map the weights correctly to avoid these complications.   
Moreover, the \ac{AIHWKIT} supports learning the digital output scales during training, either as tile-wise or column-wise scales (\longvar{out_scaling_columnwise}), which is enabled with \longvar{learn_out_scaling}.  

\subsubsection{Output Noise}
When an analog \ac{MVM} is performed, weight-independent noise from the peripheral circuits at the crossbar output is introduced, from sources such as operational transconductance amplifiers used in \acp{ADC}. This is referred to as \emph{output noise}, which is called $\sigma_\text{out}$ in \eqref{mat-vec-analog}. In the \ac{AIHWKIT}, output noise is assumed to be additive Gaussian, i.e., it is sampled from a normal distribution centered around zero. The standard deviation of the output noise $\sigma_\text{out}$ can be specified with \longvar{out_noise} (see \tabref{mvm-nonidealities}).

\subsubsection{Short-term Weight Noise}
In addition to output noise, when performing \acp{MVM}, weight-dependent noise, referred to as short-term weight noise, can be applied. In \eqref{mat-vec-analog} this noise corresponds to $\sigma_\text{w}$. This Gaussian noise of zero mean thus models variations in the weights that occur every time an \ac{MVM} is performed, such as short-term read fluctuations. For efficiency of implementation, this noise is applied on the output $y_i$, and therefore does not modify the actual weight matrix from one mini-batch to the next. In principle, the $\sigma_\text{w}$ could be function of actual conductances and inputs. The \ac{AIHWKIT} so far supports three different types of short-term weight noise, which are listed and described in \tabref{short_term_weight_noise_types}. The weight noise type is specified by \longvar{w_noise_type} and its standard deviation or scale by \longvar{w_noise} (see \tabref{mvm-nonidealities}).

\begin{table}[!t]
\centering
\begin{threeparttable}
\caption{Types of short-term weight noise set using the \longvar{w_noise_type} property.}
\label{tab:short_term_weight_noise_types}
\begin{tabularx}{\textwidth}{Xp{0.75\textwidth}}
    \toprule \toprule
    \textbf{Type} & \textbf{Description} \\ \midrule
\longvar{NONE} & Do not apply short-term weight noise.\\
\longvar{ADDITIVE_CONSTANT}                  & Apply constant additive noise with a standard deviation given by \longvar{w_noise}. Note that the weight noise is applied directly to the mapped weights (they can be accessed with \longvar{get_weights(apply_weight_scaling=False)}) which are typically in the range $-1,\ldots, 1$.                                                       \\
\longvar{PCM_READ} & Apply output-referred \ac{PCM}-like short-term read noise that scales with the amount of current generated for each output line and thus scales with both conductance values and input strength. In this case, \longvar{w_noise} specifies the scale, for which a value of $0.0175$ has been found to capture \ac{PCM} device measurements (see~\cite{rasch2023hardware} section `Short-term PCM read noise' for details).                    \\
 \bottomrule \bottomrule
\end{tabularx}
\end{threeparttable}
\end{table}

\subsubsection{Input and Output Quantization}
In conventional \ac{AIMC} systems, for each crossbar, analog-to-digital and digital-to-analog conversion is required to convert the \ac{WL} inputs and \ac{BL} outputs, using \acp{DAC} and \acp{ADC}, respectively. Due to practical constraints, these conversions are performed at a reduced precision, and thus introduce input and output quantization noise.
In the \ac{AIHWKIT}, both input and output quantization are modelled using the following assumptions: values are bounded between a fixed range, i.e., a minimum and maximum value, and $2^{N-1}$ quantization states are linearly spaced between (inclusive of) these values. 
Optionally, one can also add input and output noise to model conversion inaccuracies and S-shaped output non-linearity to model non-linear \ac{ADC} saturation.

Generally, the input (\ac{DAC}) and output (\ac{ADC}) quantization is modelled uniform quantization between symmetric bounds around zero. In more detail, it is

\begin{equation}
  \label{eq:quant}
  \quant_b^r(z) \equiv \clip{-b}{b}{2 b r\, \round\left(\frac{z}{ 2 b r}\right)},
\end{equation}
where the resolution $r$ controls the number of bins $1/r$ in the range $-b, \ldots, b$. 
The input and output resolution can be specified using the \longvar{inp_res} and \longvar{out_res} properties of the \longvar{IOParameters}, respectively, and the bounds with \longvar{inp_bound} and \longvar{out_bound}  (see~\tabref{mvm-nonidealities}). 

The resolution can either be set as the number of discrete values using an integer value, or the distance between each discrete value (the resolution), using a floating point value. Assume that the bound is set to 1 and the resolution to $1/2$. This would result in a partition of 3 bins, namely $-1 \le x < -\frac{1}{2}$,  $-\frac{1}{2} \le x < \frac{1}{2}$, and $\frac{1}{2} \le x \le 1 $ (where the value $x$ is clipped at the bounds). This would need at least 2 bits to code in digital (one of the $2^2=4$ values is discarded). Thus, in general, to set a bit resolution of e.g. $n_\text{bit}=8$ the resolution parameters need to be set to either $(2^{n_\text{bit}} - 2)$ or $1.0 / (2^{n_\text{bit}} - 2)$. If this is set to $-1$, quantization noise is not modelled, however, the clipping bound is still applied.
Stochastic rounding~\cite{Croci2022} can be modelled by enabling the boolean \longvar{inp_sto_round} and \longvar{out_sto_round} properties.

Input and output bounds, i.e., the clipping bounds/ranges for \acp{ADC} and \acp{DAC} (see \eqref{quant}), can be specified using the \longvar{inp_bound} and \longvar{out_bound} properties, respectively (see \tabref{mvm-nonidealities}).
The input bound corresponds to the maximum (read) voltage amplitude/duration for a given \ac{WL} input. Typically, we assume that the \longvar{inp_bound} is set to 1.0, so that the voltage is given in normalized units and maximally 1. To convert the actual input range into this normalized units, an additional scalar factor is used, that can also be learned (see \secref{input-range}) or dynamically set (see \secref{noise-management} for details).    

The output bound is a design choice referring to the maximally accumulated currents before the \ac{ADC} saturates. Typically we assume that weights are given in normalized units as well and clipped at maximal 1 (which needs to be ensured by enabling remapping or clipping in case of \ac{HWA} training (\secref{inference}) or is set as a material device property during training (\secref{training})). Thus, if \longvar{out_bound} is set to 10 (the default), the \ac{ADC} will saturate when more than 10 inputs are maximally on (1) while all weights are set to the maximal conductance (1). In other words, the output bound can be interpreted as corresponding to the maximum number of devices in a given column that can be at a maximum conductive state when all corresponding \ac{WL} inputs are at a maximum i.e., 1.0, and all other \ac{WL} inputs are disabled, before hard \ac{ADC} saturation occurs. 

\subsubsection{IR Drop}
Ideally, for each crossbar, the voltage along each \ac{BL} can assumed to be constant. In a real crossbar, however, finite wire resistance causes current and voltage drops between adjacent rows and columns. This phenomena is commonly referred to as IR drop~\cite{Chen2013}, and can be accurately modelled using a number of non-linear differential equations. In the \ac{AIHWKIT}, to keep the simulation-time reasonable when modelling IR drop, a number of approximations are made. Firstly, IR drop is modeled independently for each \ac{BL}, as column-to-column differences are implicitly corrected (to first order) when programming weights with an iterative-based programming scheme. Secondly, only the average integration current is considered. Lastly, the solution is approximated with a quadratic equation. We refer to \cite{rasch2023hardware} for more details. The scale of IR drop \longvar{ir_drop} and the physical ratio of wire conductance from one cell to the next to physical max conductance \longvar{ir_drop_g_ratio} can be set as part of the \longvar{IOParameters} (see \tabref{mvm-nonidealities}). The latter default value is computed with 5$\mu$S maximal conductance and 0.35~$\Omega$ wire resistance, i.e., $(1 / 0.35/ 5\cdot 10^{-6}) = 571428.57$. Note that the approximations made here to obtain a fast implementation do not allow an arbitrary setting of this parameter. The approximations only hold when the order of magnitude of this default value is not changed. 

\subsubsection{Noise and Bound Management}
\label{sec:noise-management}
To avoid the operation of peripheral circuitry in non-linear regimes, and to improve signal quality, noise and bound management can be employed~\cite{Gokmen2017}.
Noise management is used to dynamically re-scale inputs using a linear factor, $\alpha$, prior to digital-to-analog conversion to match the (fixed) input range, and bound management is used to dynamically avoid or minimize the amount of output clipping (e.g. by dynamically recomputing with down-scaled inputs when outputs were clipped). Note that while these dynamic techniques often improve accuracy, they also may implicate higher chip complexity to implement additional (\ac{FP}) operations needed, which typically translate to higher run time, energy or performance costs (not captured with \ac{AIHWKIT}).  Thus the user needs to carefully adjust these settings as appropriate for the hardware under consideration. In any case, the \ac{AIHWKIT} can readily be used to quantify the impact on accuracy when enabling such dynamic compensation methods for a given AI workload.  

Different types of noise and bound management strategies are available (see documentation of \longvar{NoiseManagementType} and \longvar{BoundManagementType}). By default, the following bound and noise management strategy types are used:
\begin{minted}[breaklines]{python}
    rpu_config.forward.bound_management = BoundManagementType.ITERATIVE
    rpu_config.forward.noise_management = NoiseManagementType.ABS_MAX
\end{minted}
The noise management type \longvar{NoiseManagementType.ABS_MAX} sets initially $\alpha \equiv max_i |x_i|$ and thus devides the input by the absolute maximum, e.g. $\mathbf{x} / \alpha$,  before reaching the \ac{DAC} and then re-scales the output of the \ac{ADC} with $\alpha$ again.  
For \longvar{BoundManagementType.ITERATIVE}, the \ac{MVM} is recomputed iternatively with setting $\alpha \leftarrow \alpha/2$ until the output bounds are not clipped anymore.  \longvar{max_bm_factor} sets the maximal bound management factor (if this factor is reached, the iterative process is stopped), and \longvar{max_bm_res} sets the maximum effective resolution number of the inputs.
It is noted that, for inference, noise/bound management is typically disabled/not used, as it requires additional computational resources to be implemented in hardware and is not supported in typical \ac{AIMC} inference chips.

\subsubsection{Other MVM Nonidealities}
In addition to the aforementioned nonidealities, the \ac{AIHWKIT} can be used to simulate many other \ac{MVM} nonidealities, including, but not limited to: voltage offset variation, device polarity read dependence, output asymmetry, and S-shaped non-linearity. We refer the reader to the API documentation of the \href{https://aihwkit.readthedocs.io/en/latest/api/aihwkit.simulator.configs.utils.html?highlight=IOParametersaihwkit.simulator.configs.utils.IOParameters#aihwkit.simulator.configs.utils.IOParameters}{\texttt{IOParameters}} for a comprehensive list of parameters and values, which have not been explicitly described in this section.

\clearpage
\section{Analog in-memory DNN inference}
\label{sec:inference}
As previous mentioned, the \ac{AIHWKIT} can be used to accurately model \ac{AIMC} \acp{MVM}, and by extension, DNN inference, by simulating a large variety of device and circuit-nonidealities. In this section, we introduce additional nonidealities used to model DNN inference. Additionally, techniques for training for inference, also referred to as \ac{HWA} training, will be discussed. We also describe how externally trained models can be imported into the \ac{AIHWKIT} to perform inference evaluation simulations and  discuss best practices for inference evaluation.

We assume that the reader is familiar with the \ac{AIHWKIT} high-level design (\secref{aihwkit}) and how to configure the hardware characteristics using the \RPUConfig\ (\secref{rpu-config}). In particular, here we discuss the situation of investigating a chip that is designed for \ac{AIMC} inference only, so that the \RPUConfig\ is derived from the \longvar{InferenceRPUConfig} class (see \tabref{rpu-configs-types}). We will discuss the additional \RPUConfig\ fields available for this case. 

\subsection{Noise Models for Inference}\label{sec:noise-inference}
In \secref{aihwkit}, configurable \ac{MVM} nonidealities are described, which can be used for modelling both \ac{DNN} on-chip inference and training. 
In the following subsections, we introduce additional deviations and long-term effects on the weights, which are specific to \ac{DNN} inference.

When evaluating a given analog model for inference accuracy, prior to the inference evaluations, programming noise as well as long-term effects up to a time $t_\text{inf}$ (such as drift and accumulated read noise, see below) need to be applied. In \ac{AIHWKIT} this is done with special methods:
\begin{minted}[breaklines]{python}
    analog_model.program_analog_weights()
    analog_model.drift_analog_weights(t_inf)
\end{minted}
Thereafter, the test set can be evaluated with the correctly applied long-term effects to the model. In the following, we describe in more detail what noise and compensations are applied during these calls. 

\subsubsection{Phenomenological Weight Noise Models}
\begin{figure}[!t]
    \centering
    \includegraphics[width=1.0\textwidth]{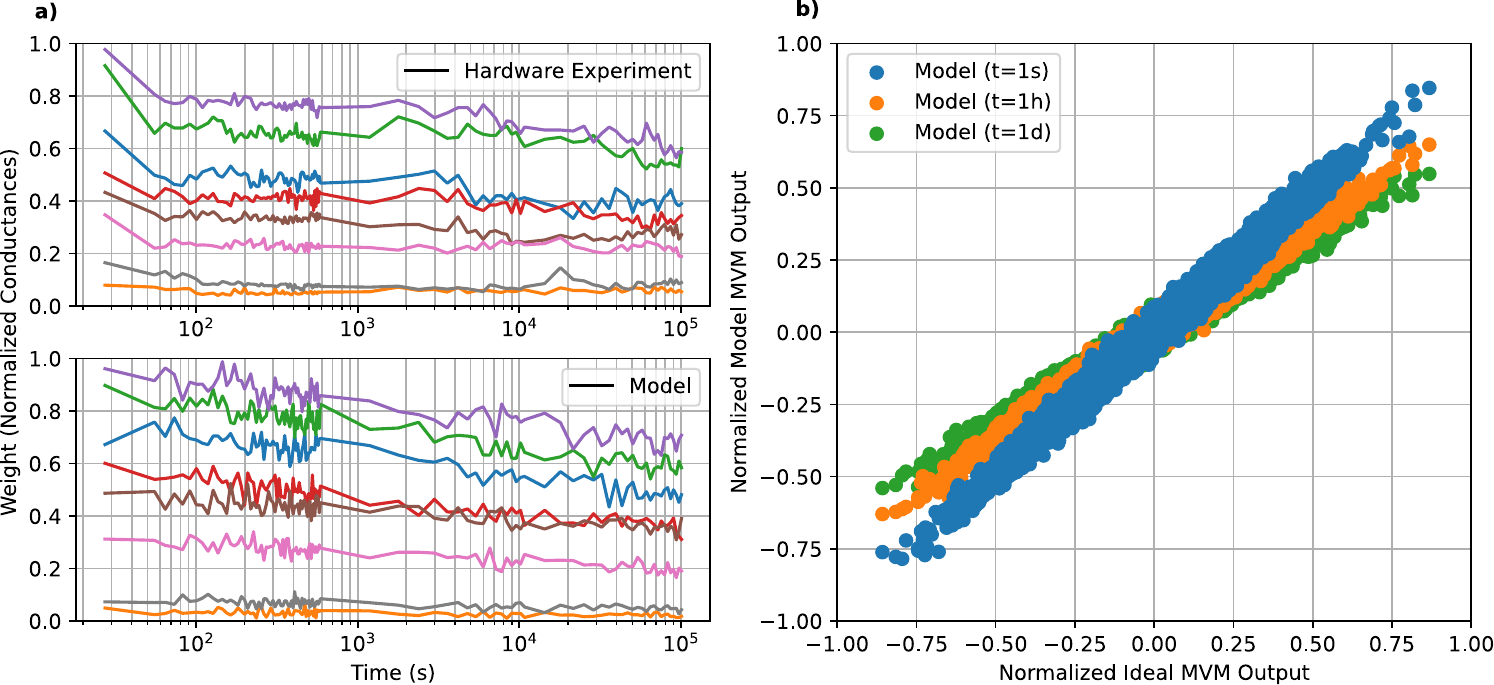}
    \caption{(a) Experimentally (hardware) obtained temporal evolution of \ac{PCM} conductance~\cite{Joshi2020} compared to that simulated by the \ac{AIHWKIT} \longvar{PCMLikeNoise} statistical noise model. Note that it is assumed all weights are programmed at the same time in the simulation, whereas in the experiment, devices converged at different iterations of programming. (b) Non-ideal \acp{MVM} from a $512\times512$ analog tile simulated using the \ac{AIHWKIT} with commonly used settings, as listed in \tabref{mvm-nonidealities}, and the \longvar{PCMLikeNoise} statistical noise model. Inputs are sampled from a sparse uniform distribution, with a sparsity of 50\%, and weights are sampled from a clipped Gaussian distribution with a standard deviation of 0.246. For $t=1$s, the reported $L_2$ error of the \ac{MVM} is 13\%.
}
    \label{fig:nonidealities}
\end{figure}

During inference, weight programming error, conductance drift, and read noise, are modelled using phenomenological noise models. Some of these models, such as the \longvar{PCMLikeNoiseModel}~\cite{nandakumarICECS} and \longvar{ReRamWan2022NoiseModel}~\cite{wan2022compute} are hardware-calibrated. The \ac{PCM} model is calibrated using a large number of device measurements, as depicted in \figref{nonidealities}. The phenomenological noise model to use can be specified using the \longvar{noise_model} field of the \RPUConfig, as follows:
\begin{minted}[breaklines]{python}
    from aihwkit.inference import PCMLikeNoiseModel
    rpu_config.noise_model = PCMLikeNoiseModel()
\end{minted}
Note that most inference-only related classes and tools can be imported from \longvar{aihwkit.inference}. 

\paragraph{Weight Programming Error}
When programming real \ac{NVM} devices, the programmed conductances, $g^{\text{P}}_{ij}$, differ from the desired target values, $\hat{g}_{ij}$, due to many underlying mechanisms, including, but not limited to: cycle-to-cycle and device-to-device variability, \ac{WL} and \ac{BL} voltage mismatches, device-level voltage asymmetries~\cite{ghazi2022}, and temporal drift. While many of these mechanisms can be emulated for a given programming scheme to infer the weight programming error, it is much more computationally efficient to compute the programming error using an arbitrary function, $g^{\text{P}}_{ij} = f(\hat{g}_{ij})$, which is defined for each device model and programming scheme. It is typically assumed that the weight error can be modelled using a normal distribution centered around $\hat{g}_{ij}$, where the standard deviation, $\sigma$, is dependent on $\hat{g}_{ij}$, as follows:
\begin{equation}\label{eq:prog-noise}
    g^{\text{P}}_{ij} = {\cal N}\left(\hat{g}_{ij}, \sigma(\hat{g}_{ij})\right).
\end{equation}

In the \ac{AIHWKIT}, the \longvar{apply_programming_noise_to_conductance(g_target)} method of the noise model (base class) is used to apply the weight programming error. For more details and how to customize the noise model see \secref{extension}.

\paragraph{Conductance Drift}
Many types of \ac{NVM} devices, most prominently, \ac{PCM}, exhibit temporal evolution of the conductance values, referred to as the conductance drift. This poses challenges for maintaining synaptic weights reliably~\cite{Boybat2018}. Conductance drift is most commonly modelled using \eqref{drift}, as follows:
\begin{equation}\label{eq:drift}
    g(t) = g(t_0)(t/t_0)^{-\nu},
\end{equation}
where $g(t_0)$ is the conductance at time $t_0$ and $\nu$ is the drift exponent. In practice, conductance drift is highly stochastic because $\nu$ depends on the programmed conductance state and varies across devices. In the \ac{AIHWKIT}, the \longvar{apply_drift_noise_to_conductance(g_prog, nu_drift, t_inference)} method of the noise model (base class) is used to apply the conductance drift noise.

\paragraph{Low-Frequency Read Noise}
When devices are read, after the conductances have been programmed, there will be instantaneous fluctuations on the hardware conductances due to the intrinsic noise from the \ac{NVM} devices. Many \ac{NVM} devices exhibit $1/f$ noise and random telegraph noise characteristics, which alter the effective conductance values used for computation. This noise is referred to as read noise, because it occurs when the devices are read after they have been programmed. Note that here we refer to longer-term and lasting effects on the conductances after programming such as low-frequency 1/f fluctuations (typically much slower than processing a single mini-batch) as opposed to weight read fluctuations on the time-scale of a single \ac{MVM}. Therefore, this read noise is resampled only once at every inference time $t_\text{inf}$. Short-term read fluctuations that are resampled every \ac{MVM} can be instead set using the \longvar{IOParameters} as listed in \tabref{mvm-nonidealities}.

The low-frequency read noise is typically modelled using a normal distribution centered around zero with a standard deviation of $\sigma_{nG}$ dependent on the time elapsed since programming, i.e., ${\cal N}\left(0, \sigma_{nG}(t)\right)$~\cite{nandakumarICECS}. The conductance of device a function of time, accounting for both conductance drift and read noise, can be modelled using \eqref{conductance-time}, as follows:
\begin{equation}\label{eq:conductance-time}
    g(t)= g_\text{drift} (t)+ {\cal N}\left(0,\, \sigma_{nG}(t)\right).
\end{equation}
In the \ac{AIHWKIT}, the \longvar{apply_noise(weights, t_inference)} method of the noise model (base class) is used to apply both conductance drift and read noise.

\subsubsection{Drift Compensation}
Various methods can be employed to mitigate the effect of conductance drift  during inference~\cite{2019AmbrogioIEDM}. In the \ac{AIHWKIT}, such techniques are referred to as \emph{drift compensation techniques}. As proposed in~\cite{legalloTED2018}, a single scaling factor, $\gamma$, can be applied to the output of an entire crossbar (after the \ac{ADC}) in order to compensate for a global conductance shift. In the \ac{AIHWKIT}, to compute the correct value for a time $t_\text{inf}$ after the conductance programming (at $t_\text{inf}=0$), first a measure for the strength of a reference output using \acp{MVM} right after programming is stored in $s_0$. When compensating after a time $t_\text{inf}$, the same \acp{MVM} are computed with the drifted weights to get another output strength $s_t$. The compensation factor is then set to $\gamma \equiv s_0 / s_t$. For the global drift compensation (\longvar{GlobalDriftCompensation}) the output strength is computed as the mean absolute $y_i$ values resulting from giving all one-hot vectors as input. However, other strength measures can be implemented by customizing the drift compensation as explained in \secref{extension}.   

The drift compensation type can be specified using the \longvar{drift_compensation} field of the \RPUConfig:
\begin{minted}[breaklines]{python}
    from aihwkit.inference import GlobalDriftCompensation
    rpu_config.drift_compensation = GlobalDriftCompensation()
\end{minted}

\subsection{Hardware-aware Training for Inference}
\ac{HWA} training, a popular alternative to on-chip training, can also be used to train networks for deployment on \ac{AIMC} hardware. Unlike on-chip training, \ac{HWA} training is solely performed in software, and does not require detailed behavioural or physical device models. Instead, additional operations, such as weight noise injection, are added during forward and backwards propagation passes, and standard SGD methods are used. These are added to increase the model robustness~ \cite{tsai2019inference, Joshi2020,  yang2022tolerating, gokmen2019marriage, Kariyappa2021, Spoon2021,rasch2023hardware}, and can be specified using different \RPUConfig\ parameters (as part of the \longvar{InferenceRPUConfig} class), which are discussed in the following subsections.

\subsubsection{AIMC Forward Pass During HWA Training}
It is common for \ac{HWA} training to assume a perfect backward pass, with non-idealities only added during the forward pass, which is the default behavior of \longvar{InferenceRPUConfig}. \ac{MVM} nonidealities added to the \longvar{forward} field (see \tabref{mvm-nonidealities}) of the class are applied when the model is in \longvar{train()} mode and \longvar{eval()} mode. One can configure additional noise sources that are only present when the model is in \longvar{train()} mode (see the next sections for details). While \longvar{InferenceRPUConfig} uses a C++/CUDA backend, \longvar{TorchInferenceRPUConfig} is purely based on PyTorch, making debugging easier as one is able to step through every part of the forward pass. Switching to the PyTorch based tile is as simple as exchanging \longvar{InferenceRPUConfig} with \longvar{TorchInferenceRPUConfig} (see \tabref{rpu-configs-types}).

\subsubsection{Weight Modifier Parameter}
Weight modifier parameters (\longvar{WeightModifierParameter}), set using the special field \longvar{modifier} of the \RPUConfig, are used to specify different attributes about the injected weight noise during \ac{HWA} training, such as the noise type and amplitude. In \tabref{type_weight_noise}, a description of each weight modifier parameter type is provided. When a weight modifier type other than \longvar{COPY} is used, unless otherwise specified, for the duration of a mini-batch, each weight will be modified during both forward and backward propagation cycles.
Drop connect~\cite{pmlr-v28-wan13,gokmen2019marriage,li2023impact}, which is used to set weights to zero with a given probability during training, can be used with any other modifier type in combination. 
As an example, additive Gaussian noise with a standard deviation of 0.1 can be applied, in addition to drop connect, with a drop connect probability of 0.05, as follows:
\begin{minted}[breaklines]{python}
    from aihwkit.simulator.configs import WeightModifierType
    rpu_config.modifier.type = WeightModifierType.ADD_NORMAL
    rpu_config.modifier.std_dev = 0.1
    rpu_config.modifier.pdrop = 0.05
\end{minted}

For relatively small networks and datasets, we found that increasing the number of times we draw samples from our weight distribution improves the robustness to programming noise of our model. This can be achieved by adding noise drawn from the distribution specified by \longvar{WeightModifierType} for every sample in the batch. Concretely, for inputs of shape \longvar{[batch_size,d_in]} and a layer weight of shape \longvar{[d_in,d_out]}, instead of applying noise to the weights once, yielding again a matrix of shape \longvar{[d_in,d_out]}, we add noise for every sample in the batch, yielding a weight matrix of shape \longvar{[batch_size,d_in,d_out]}. This feature can be turned on by setting \longvar{rpu_config.modifier.per_batch_sample} to \longvar{True}. Note that this feature is only available for the \pytorch-based analog tile implementation, which can be selected by using \longvar{TorchInferenceRPUConfig} as the \longvar{rpu_config} class.

\begin{table}[!t]
\centering
\begin{threeparttable}
\caption{Types of weight modifiers. Some experimental weight modifier types, including \longvar{WeightModifierType.DOREFA}, are not listed. Parameters are grouped in the class \longvar{WeightModifierParameter} and accessible in the \longvar{modifier} attribute of the \RPUConfig.}
\label{tab:type_weight_noise}
\begin{tabularx}{\textwidth}{Xp{0.75\textwidth}}
    \toprule \toprule
\textbf{Type} & \textbf{Description} \\ \midrule
\longvar{NONE} & No weight modifier is applied. \\
\longvar{DISCRETIZE} &  Weights are discretized (quantized) according to the resolution specified by \longvar{res}. If \longvar{sto_round} is enabled, stochastic rounding is performed.\\
\longvar{MULT\_NORMAL} & Multiplicative Gaussian noise is added to all weights with a standard deviation of \longvar{std_dev}.\\
\longvar{ADD\_NORMAL} & Additive Gaussian noise is added to all weights with a standard deviation of \longvar{std_dev}.\\
\longvar{POLY} & Noise is added to all weights from a normal distribution with a standard deviation of $\sigma_\text{wnoise} (c_0 + c_1 |w_{ij}|/\omega + c_N |w_{ij}|^N/\omega^N)$, where $\omega$ is either the actual absolute max weight (if \longvar{rel_to_actual_wmax} is set) or the value \longvar{assumed_wmax}. $\sigma_\text{wnoise}$ is set using the \longvar{std_dev} parameter. The coefficients $c_0, ..., c_N$ are set using the \longvar{coeffs} parameter. \\
\longvar{PROG_NOISE} & Identical to \longvar{POLY} except that a positive or negative weight will remain positive or negative, respectively, after the noise is applied to simulate the situation of programming the weight to two separate conductances depending on the sign. If weights change sign after applying noise, the absolute value with preserved sign is taken.   \\
\bottomrule \bottomrule
\end{tabularx}
\end{threeparttable}
\end{table}

\subsubsection{Weight Clipping and Remapping Parameter}
Weight clipping and remapping ensures that the weight is correctly mapped to (normalized) conductances in the range and thus should always be applied during \ac{HWA} training (at least \longvar{fixed_value} clipping to 1) to avoid that unrealitstic weight ranges that are not in line with the assumptions when specifying the other MVM nonidealities (such as \ac{ADC} range etc., see \tabref{mvm-nonidealities}). Note that the weight range here refers to the \emph{analog weight} $\analog{W}$. The actual \ac{FP} weight is given by the $\analog{W}$ times the (digital) output scaling parameters (see \secref{mapping} for details).

Weight clipping parameters (\longvar{WeightClipParameter}), set using the special field \longvar{clip} of the \RPUConfig, are used to specify different attributes that control how weights are clipped during \ac{HWA} training. In \tabref{type_clip_remap}, different weight clipping technique types are listed.
Weight remapping parameters (\longvar{WeightRemapParameter}), set using the special field  \longvar{remap} of the \RPUConfig, are used to specify different attributes that control how weights are re-mapped to analog weights $\analog{W}$ and the output scales $\alpha^\text{out}$ during \ac{HWA} training using the assumption of having digital output scales that can represent part of the full weight together with the value represented in the conductances (see \secref{mapping}). The \longvar{remapped_wmax} parameter specifies the assumed maximum analog weight value. This is typically set to 1.0.
In \tabref{type_clip_remap}, different weight remapping parameters are listed.
As an example, weight clipping using \longvar{LAYER_GAUSSIAN} at 2 times the standard deviation of the weight distribution, and weight remapping (in \longvar{CHANNELWISE_SYMMETRIC} mode) can be enabled as follows:
\begin{minted}[breaklines]{python}
    from aihwkit.simulator.configs import WeightClipType, WeightRemapType
    rpu_config.clip.type = WeightClipType.LAYER_GAUSSIAN
    rpu_config.clip.sigma = 2.0
    rpu_config.clip.fixed_value = 1.0
    rpu_config.mapping.type = WeightRemapType.CHANNELWISE_SYMMETRIC
\end{minted}
Note that mapped weights in the analog representation should always be smaller than the assumed maximal value (typically 1), to ensure that this clipping at fixed value can be used in combination.  

\begin{table}[!t]
\centering
\caption{Types of weight clipping and remapping techniques.}
\label{tab:type_clip_remap}
\begin{tabularx}{\textwidth}{Xp{0.75\textwidth}}
\toprule \toprule
\textbf{Type} & \textbf{Description} \\ \midrule
\longvar{NONE} & Clipping/remapping behaviour is disabled. \\ \midrule
\multicolumn{2}{c}{\textbf{Weight Clipping}} \\ \midrule
\longvar{FIXED\_VALUE} & Weights are clipped to fixed value give, symmetrical around zero, specified by \longvar{rpu_config.clip.fixed_value}.\\
\longvar{LAYER\_GAUSSIAN} & Calculates the second moment of the whole weight matrix and clips at $\sigma$ times the result symmetrically around zero. $\sigma$ is specified using the \longvar{rpu_config.clip.sigma} parameter.\\
\midrule
\multicolumn{2}{c}{\textbf{Weight Remapping}} \\ \midrule
\longvar{LAYERWISE\_SYMMETRIC} & Remap according to the absolute max of the full weight matrix.\\
\longvar{CHANNELWISE\_SYMMETRIC} & Remap each column (output channel) in respect to the absolute max.\\
\midrule \midrule
\end{tabularx}
\end{table}

\subsubsection{Setting and Learning the Input Ranges}\label{sec:input-range}

As previously described, inputs are first clipped to a fixed range before being presented to each crossbar. The input range for each crossbar can either be learned during training, dynamically computed during inference, or fixed (set manually). 

In the \ac{AIHWKIT}, pre-post processing parameters, specified using \longvar{PrePostProcessingParameter}, can be used to augment digital input and output processing steps. Currently, input range learning is the only natively supported processing step. Input range learning can be used to find the optimal input range for each crossbar during \ac{HWA} training. For initialization, one can use the first \longvar{init_from_data} input batches for calculating a moving average of the \longvar{init_std_alpha}$^{th}$ standard deviation of the input distribution. After the amount of batches have been presented, learning takes over. This is done by calculating the gradient of the input range to be proportional to the amount of clipping caused by the current input range and the gradient of the crossbar inputs. This typically widens the input range so that no clipping occurs, however, a tight input range is often more favorable since it reduces quantization error and boosts the overall signal strength, which is important if the hardware suffers from output noise. How much the input range is tightened at every backward pass can be controlled via the \longvar{decay} attribute which adds \longvar{input_range * decay} to the gradient if not more than some percentage of the inputs is clipping. This percentage can be controlled via \longvar{input_min_percentage}. As an example, using a value of 0.95 as \longvar{input_min_percentage} will only lead to a tightening of the input range if less than 5\% of the inputs have been clipped using the current input range. The input range can also be loosened up if the outputs are clipping at the \ac{ADC}. This can be turned on by setting \longvar{manage_output_clipping=True}. Again, for \longvar{output_min_percentage=0.95}, the input range is not loosened if less than 5\% of the outputs are clipping. It should be noted that this feature is currently not supported in the torch-based tile. By default, the gradient of the input range (before decaying) is scaled by the current input range. To turn this feature off, set \longvar{gradient_relative=False}. For an example on how to use input range learning, see notebook \texttt{hw\_aware\_training.ipynb}\cite{hw_aware_training}.

If learning the input ranges is not desired, but \ac{HWA} training with \acp{DAC} and \acp{ADC} is, then a second option is to use the \longvar{NoiseManagementType} to dynamically scale the inputs during \ac{HWA} training and inference such that each input covers the full input range. However, note that this is typically not supported by most \ac{AIMC} hardware due to the high computational overhead involved in implementing this dynamic range computation. For more details, refer to \secref{noise-management}.

To simplify the \ac{HWA} training, one might eventually want to train without \acp{DAC} and \acp{ADC} altogether, in which case, one can simply enable a perfect forward pass by setting \longvar{forward.is_perfect = True} in the \RPUConfig. In this case, one has to calibrate them post-training before deploying them on hardware. Setting the input ranges post-training typically involves calibration using a subset of the training data. During the calibration phase, the model is in evaluation mode, which means that layers such as \longvar{torch.nn.Dropout} operate in inference mode, and any distortions such as output noise, weight noise, or input quantization are turned off. The activations from every crossbar are then cached until no more inputs are provided. To avoid exhausting the memory, one can set an upper limit of activation samples cached at every crossbar. In order to prevent sampling of activations that are not representative of the true distributions, new samples are randomly mixed into the cache, which is then trimmed to the maximum number of samples. After the sampling phase, the \longvar{input_range} field of every crossbar is populated with a certain quantile of the recorded samples. This ensures that outliers are not mapped to the full range causing an overall weak signal for the intermediate values. This mode is demonstrated in notebook \texttt{post\_training\_input\_range\_calibration.ipynb}\cite{calibration}.

For large models, caching even a couple of hundred activation samples per crossbar might already be too memory intensive. For this reason, a moving average of the quantile can be computed. This drastically reduces the memory footprint since no caching is required, but still enables input range calibration on large amounts of data. However, the moving average is of course an approximation to the true quantile, which might lead to worse performance.

\subsubsection{Importing Externally Trained Models}
Externally trained models can be be imported to the \ac{AIHWKIT}, either to be retrained using \ac{HWA} training for inference, or for direct inference evaluation. Currently, the \ac{AIHWKIT} natively supports conversion of \pytorch\ models, so models trained using other \ac{ML} frameworks first require conversion to a \pytorch-based model. External libraries, such as those listed here\footnote{\url{https://github.com/ysh329/deep-learning-model-convertor}}, can be used to convert trained models from many popular libraries to \pytorch-based models.

All linear (dense) and convolutional layers of an arbitrary \pytorch-based model can be automatically converted to analog equivalent layers using the \longvar{aihwkit.nn.conversion.convert_to_analog(module, rpu_config)} methods, where the \ac{AIMC} hardware properties (including tile size etc.) are defined in \longvar{rpu_config}. Other layers, namely \ac{LSTM} cells, require manual in-place conversion.

It should be noted that most imported models do not have pre-calibrated input ranges, which is why, most of the times, one needs to calibrate them after loading the model. For information on how to do that, see the previous section.

\subsubsection{Hardware-aware Training Example}
For \ac{HWA} training one typically starts off from a model that was pre-trained without any \ac{AIMC} nonidealities or techniques such as weight clipping or noise injection. If there is a need for training from scratch, the user can either define the network in \pytorch\ and then convert it using \longvar{convert_to_analog} or directly substitute the individual layers with their analog counterparts in the model definition. Notebook \texttt{hw\_aware\_training.ipynb}\cite{hw_aware_training} demonstrates this workflow with a ResNet-32 trained on the Cifar-10~\cite{cifar10} dataset. We start off by pre-training the model to the baseline accuracy, which in this case hovers around $94\%$. For the \ac{HWA} training, we first generate an \RPUConfig\ that then is used when converting the model to analog. For training an analog network, one has to use an \longvar{AnalogOptimizer}, which adds specific logic to be executed after parameter updates. In this case, we use the simple \longvar{AnalogSGD}, however more complex algorithms can be used by mixing \longvar{AnalogOptimizerMixin} into the \pytorch-based optimizer class (see \longvar{AnalogAdam} for an example). It should be noted that for \ac{HWA} training, the learning rate might need to be reduced. By how much depends on the network, but reducing it by roughly one order of magnitude is a good starting point. Apart from that, we are able to use the same training code to do \ac{HWA} training on the converted analog model, since all \ac{HWA} training parameter are automatically applied as defined in the \RPUConfig. After \ac{HWA} training, we perform inference using the  the model, which is now in \longvar{eval()} mode (see the next section for more information). 

\subsection{Inference Accuracy Evaluation}
For a given \RPUConfig\, during inference evaluation of the analog model, the parameter setting specific to the \ac{HWA} training, such as the specified weight noise modifier type, i.e., \longvar{rpu_config.modifier.type}, is not used (unless \longvar{modifier.enable_during_test} is explicitly set to \longvar{True} for debugging purposes). The \ac{MVM} non-idealities, as specified by \longvar{forward} field of the \RPUConfig\ (see \tabref{mvm-nonidealities}) are, however, \emph{always} applied (during \ac{HWA} training as well as inference evaluation), since they define the \ac{AIMC} properties rather than any extra regularization techniques for \ac{HWA} training. 
As described in more detail in \secref{noise-inference}, programming noise can be applied by calling either the \longvar{analog_model.program_analog_weights()} or \longvar{analog_model.drift_analog_weights(t_inference)} methods, which both inject programming noise using the \longvar{rpu_config.noise_model}. For the latter, in addition to programming noise, the current reference weights (i.e., the conductance state of all devices) are drifted for \longvar{t_inference} seconds.

\subsubsection{Multiple Model and Evaluation Instances}
As \ac{AIMC} hardware is inherently stochastic, a single evaluation instance is typically not representative of the behaviour of the modeled hardware over multiple evaluation instances. Consequently, multiple evaluation instances should be used to evaluate both the mean and variance (typically the standard deviation) of the metrics being evaluated. Moreover, as many analog \ac{NVM} devices, such as \ac{PCM}, are susceptible to temporal conductance drift, and the behaviour of analog \ac{IMC} hardware can evolve over time, performance-based metrics for analog \ac{IMC} hardware are typically reported for a specific length of time, with respect to a reference point-in-time. This is typically defined as the point-in-time when all devices have been programmed. Ideally, multiple model (random initialization) instances should also be used.

\subsubsection{Inference Evaluation Example}
Notebook \texttt{hw\_aware\_training.ipynb}\cite{hw_aware_training} provides an inference configuration example, which uses the \longvar{PCMLikeNoiseModel} during inference. The mean and standard deviation of the test set is reported for different logarithmically-spaced time steps, from \longvar{t_inference} = 60.0s up to one year ($365 \cdot 24 \cdot 60 \cdot 60$s). For each point in time, the mean and standard deviation of the test set accuracy is reported across 5 evaluation instances. Note that we kept the number of repetitions low for this example. In practice, one should repeat the same measurements at least 10 times (we typically use 25). Soundness of the experiments can be even further improved, if computational resources allow, by training the same network multiple times and reporting the performance metrics averaged across the different model instances.

\clearpage
\section{Analog In-Memory DNN Training}
\label{sec:training}

While using \ac{AIMC} chips dedicated for inference only is a common application for in-memory acceleration,  the training of today's ever-increasing DNNs would benefit greatly from hardware acceleration as well. For that purpose, analog in-memory training algorithms have been developed (as introduced in \secref{analogai}). From the algorithmic as well as chip architecture perspective, analog in-memory training is far more challenging than solely \ac{AIMC} inference. In particular,  for in-memory SGD training, the backward pass as well the incremental update are done in-memory, and thus subject to additional noise sources and nonidealities. For the development of robust \ac{AIMC} training algorithms, it is thus especially important to have good estimates of attainable accuracy assuming a particular device material, as well as being able to determine the limits of device material properties that still guarantee convergence of the training algorithm.            

The \ac{AIHWKIT} provides a particularly rich set of tools for the testing and development of \ac{AIMC} training algorithms. Out-of-the box, it provides na\"ive in-memory \ac{SGD} using stochastic pulse trains~\cite{gokmen2016pulsed}, as well as improved in-memory training algorithms, such Mixed-precision\cite{nandakumar2020}, Tiki-taka I \& II~\cite{gokmen2020, gokmen2021}, as well as newest state-of-the-art algorithmic developments, namely \acf{cTTv2} and \ac{AGAD}~\cite{rasch2023fast} (see \tabref{compounds}).

\begin{table}[b]
    \centering
    \begin{tabular}{lcp{10cm}}
         Compounds & Algorithm & Update\\
         \midrule 
           {\it Vector} & In-memory SGD & $\pulsed \analog{W}$ w/ multiple devices per crosspoint   \\
        {\it MixedPrecision} & Mixed-precision \cite{nandakumar2020} &  Digital rank-update onto $\chi$, (row-wise) pulsed transfer $\chi \rowpulsed \analog{W}$\\
        {\it Transfer} & Tiki-taka \cite{gokmen2020}&  $\pulsed\analog{A}$, slow (row-wise) transfer $\analog{A}\rowpulsed \analog{W}$\\ 
        {\it BufferedTransfer} & TTv2 \cite{gokmen2021}& $\pulsed \analog{A}\rightarrow H \rowpulsed \analog{W}$, with digital matrix $H$\\  
        {\it ChoppedTransfer} & Chopped-TTv2  \cite{rasch2023fast}&  $\choppulsed \analog{A}$ with chopper, $\analog{A}\rightarrow H \rowpulsed \analog{W}$\\  
        {\it DynamicTransfer} & AGAD \cite{rasch2023fast}&  $\choppulsed \analog{A}\rightarrow H$ with dynamic offset correction, $H\rowpulsed \analog{W}$\\  
    \end{tabular}
    \caption{Compounds are derived from \texttt{UnitCell} and used to define a specialized updated behaviour of the \texttt{UnitCellRPUConfig} (e.g. set to the \texttt{device} field). To indicate a weight matrix $W$ thought of stored on an analog crossbar, we here write $\analog{W}$. To indicate a pulsed outer product update (according to \cite{gokmen2016pulsed}), we write $\pulsed$. Slow (row-wise) read and pulsed update is indicated with $\rowpulsed$ and a column-wise read (that is an \ac{AIMC} \ac{MVM} forward pass with one-hot inputs and addition to a digital matrix) is indicated with $\rightarrow$.   Note that each of the compounds has itself a number of configuration settings for exploring the hyperparameters of the optimizers.}
    \label{tab:compounds}
\end{table}

\begin{table}[h]
    \centering
    \begin{tabular}{llp{7cm}}
         Device config &   Simplified mathematical model & Functionality\\
         \midrule 
         {\it ConstantStep} & $w \leftarrow \text{clip}(w \pm \delta)$ & Update independent of current weight (conductance)\\
        {\it LinearStep} &  $w  \leftarrow  \text{clip}\left(w \pm \delta \left(1 - \gamma w\right)\right)$
& Gradual saturation towards weight bounds with clipping\\
 {\it SoftBounds} &  $w  \leftarrow  w \pm \delta \left(1 - \frac{w}{b_\pm}\right)$ & Gradual saturation towards the bounds\\
        {\it PowStep} &         $w  \leftarrow  w \pm \delta \,\left(\frac{b_\pm - w}{b_{+} -
    b_{-}}\right)^{\gamma}$
 & Power dependency on weight\\
        {\it ExpStep} & $w  \leftarrow  w \pm  \delta \left(1 - c_0 e^{-c_1 w}\right)$ & Exponential dependency with current weight with parameters $c_0$ and $c_1$\\
        {\it PiecewiseStep} & $ w \leftarrow w \pm \delta \left((1 - q) v_k
        + q \, v_{k+1}\right)$ & User-defined nodes $v_k$ with linear interpolation, $w \in [v_k, v_{k+1}]$, $q = \frac{w - v_k}{v_{k+1} - v_{k}}$ \\
        {\it ReRamES} & Based on {\it ExpStep} & Preset setting for ReRAM\cite{gong2018signal} \\ 
        {\it ReRamArrayOM} & Based on {\it SoftBoundsReference} & Preset setting from Optimized Material ReRAM Arrays\cite{gong2022deep} \\ 
        {\it ReRamArrayHf2O} & Based on {\it SoftBoundsReference} & Preset setting from HfO2 ReRAM Arrays\cite{gong2022deep} \\ 
        {\it Capacitor} & Based on {\it LinearStep} & Preset setting for CMOS\cite{li2018capacitor} \\ 
        {\it EcRam} & Based on {\it LinearStep} & Preset setting for ECRAM\cite{tang2018}\\ {\it EcRamMO} & Based on {\it LinearStep} & Preset setting for single metal-oxide ECRAM\cite{kim2019metal}\\
        {\it GokmenVlasov} & Based on {\it ConstantStep} & Device setting used by Gokmen and Vlasov\cite{gokmen2016pulsed} \\
        {\it PCM} & Based on {\it ExpStep} and {\it OneSided} & PCM preset device pair with one-sided update (and occasional reset) \\  
    \end{tabular}
    \caption{A selection of (predefined) device models and configurations (the \texttt{Device} and \texttt{DevicePreset} name suffixes are omitted here). For \ac{AIMC} training, the update behaviour (the weight change in response to a pair of coincident voltage pulses from \ac{BL} and \ac{WL}) is governed by the \texttt{device} field of the \RPUConfig. \ac{AIHWKIT} provides numerous functional device models as well as presets, where the parameter of the functional device models are set according to measurements. All devices additionally implement device-to-device variations, where each device in the array will be set to slightly varying parameters (typically drawn from a Gaussian around the mean with user-defined variance). For instance, the actual update $\delta$ is typical computed as $\delta_{ij}  + \sigma \xi$, where $\sigma$ is the pulse-to-pulse standard deviation ($\xi\in {\cal N}(0, 1)$) and  $\delta_{ij} = \delta w_\text{min} + \sigma_\text{d-to-d} \xi$ is set at the device array construction time to model device-to-device variations (indices and details are omitted in the simplified equations above). The device is modelled in normalized weight units assuming a linear mapping of weights to conductances. For more complete equations and details see the \href{https://aihwkit.readthedocs.io/en/latest/api/aihwkit.simulator.configs.devices.html}{API documentation}}
    \label{tab:device-configs}
\end{table}

\begin{table}[h]
    \centering
    \begin{tabular}{lcp{8cm}}
         Update field & Default value &  Functionality\\
         \midrule 
         
\longvar{desired_bl} &  31 & Desired length of the pulse trains. in case of using the update BL management, it is the maximal pulse train length.\\

\longvar{pulse_type} & \textit{StochasticCompressed} &  Pulse types used when computing the outer product. Can be stochastic or implicitly deterministic. \\  
\longvar{update_bl_management} & True & Dynamic selection of the length of the pulse train as described in \cite{Gokmen2017} and \cite{rasch2023fast}.\\
\longvar{update_management} & True & Scaling of the update pulse probability to load-balance the word and bit-lines. See \cite{Gokmen2017} for details.   \\ 
\longvar{x/d_res_implicit} &  0.0 &   Resolution (ie. bin width) of each quantization step for activation $x$ or the error $d$, respectively, in case of \textit{DeterministicImplicit} pulse trains. \\
    \end{tabular}
    \caption{A selection of the possible configuration of the update pulse behavior. The most often used parameter is the desired bit length (maximal number of pulses per update) which effectively determines the maximal change of the conductance (weight). Since each pulse given to the device is of equal minimal amplitude (with possible variations determined by the device model setting), the maximal amount that can be written onto the device is the number of pulses given per update times the average change of the conductance per pulse.  Thus device update will clip at some point if the SGD demands a too large gradient update.  Small update values (smaller than the minimal update) are effectively implemented by stochastic pulsing probability smaller than one. To determine the probability the average expected minimal conductance response at (logical) zero point is used and expected to be known (\longvar{dw_min} in many device models). See \cite{gokmen2016pulsed} for details.}
    \label{tab:update-parameters}
\end{table}

\subsection{Configuration of Material Properties for In-memory Analog Training}
\label{sec:config_training}
For in-memory training, apart from the actual \ac{AIMC} training algorithm, the device material response properties are important. The fully in-memory training algorithm will typically use stochastic pulse trains and cross-point pulse coincidences to implement the outer product~\cite{gokmen2016pulsed}, or might use an intermediate digital computation before updating the analog weights matrix with incremental pulses~\cite{nandakumar2020}. In \ac{AIHWKIT}, each single incremental pulse update is explicitly modelled according to a device response model. \ac{AIHWKIT} provides highly optimized and self-tuned GPU routines to enable larger-scale \ac{AIMC} in-memory  training simulations on this level of detail. This is different from the approach for inference \secref{inference}, where statistical weight programming noise models are used instead. 

\begin{figure}
    \centering
    \includegraphics[width=\textwidth, clip, trim=5cm 1.6cm 4cm 1.6cm]{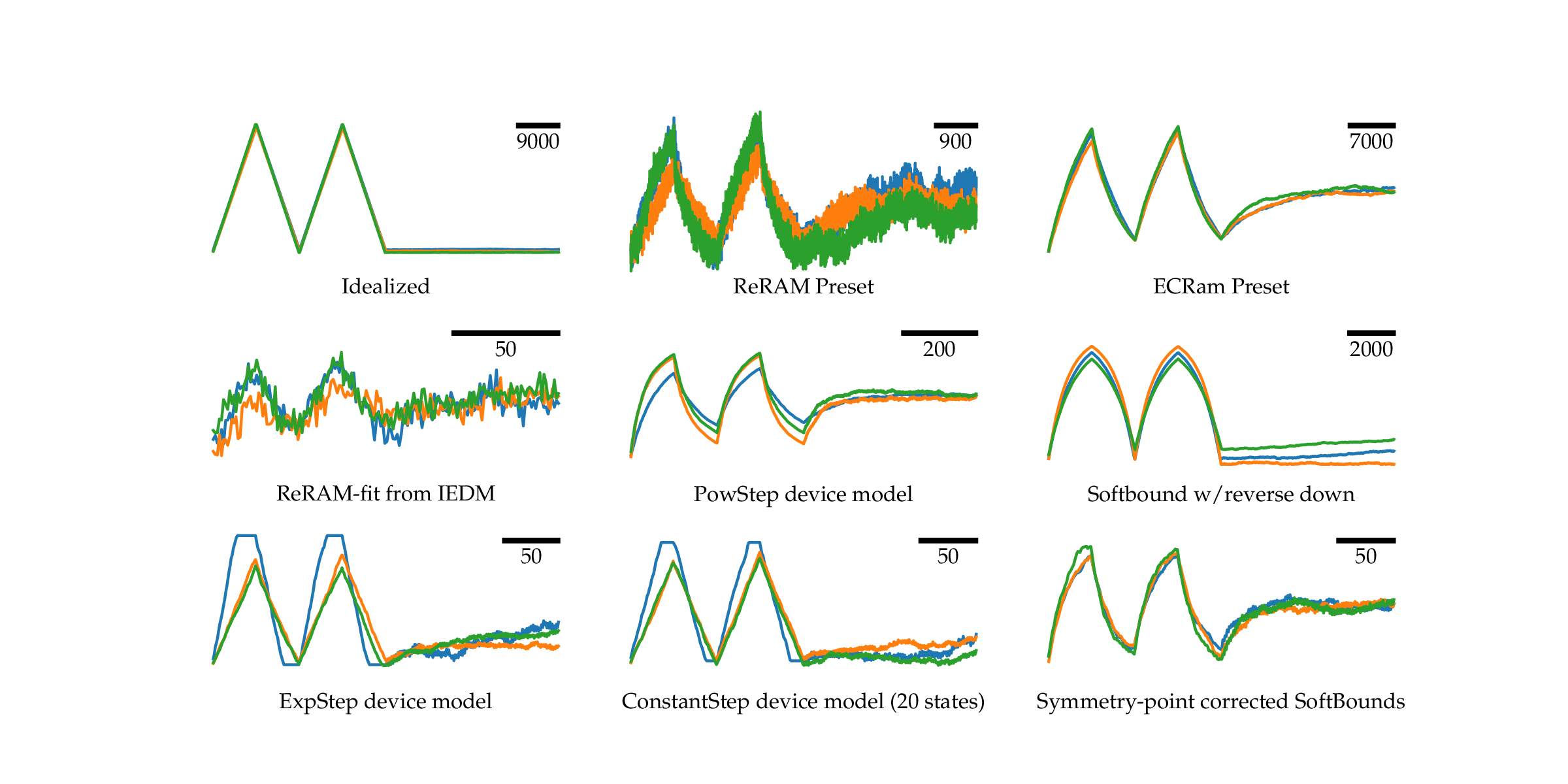}
    \caption{Example conductance responses to a series of up, down, and up-down pulses for different device configuration as listed in \tabref{device-configs}. Note that various asymmetric shapes and device-to-device variations (different colors) can be set by the user. Presets that fit measurements are available, and a fitting tool for new device measurements is provided as well. Bar shows the x-axis scale in number of pulses given. }
    \label{fig:device-plots}
\end{figure}

Material response properties for in-memory training are captured in functional device models, such as the soft-bounds model that has been used to model conductance responses to voltage pulses for \ac{RRAM} devices\cite{gong2018signal}. \ac{AIHWKIT} also provides other models and data-calibrated preset settings that are (partly) summarized in \tabref{device-configs}. See also \figref{device-plots} for an illustration.  

When setting up a \RPUConfig\ to specify an in-memory training simulation, the \emph{device configurations} listed in \tabref{device-configs} can be assigned to the \longvar{device} field or as part of a \emph{device compound}. If one wants to define a plain in-memory \ac{SGD} using stochastic pulsing~\cite{gokmen2016pulsed}, the device configuration is directly applied to the \longvar{device} field and additional properties for the update behaviour such as pulsing schemes and corrective methods are set in the \longvar{update} field~(see \tabref{update-parameters}).

However, the \longvar{device} field can also be a \emph{device compound}, in which case multiple crossbars (or parts of the more complicated unit cell at each crosspoint) are simulate according to the definition of the training algorithm~(compare to \figref{inference_training}~b-d). The specialized \ac{AIMC} update algorithms supported are listed in~\tabref{compounds}. 

In \secref{training_fit}, we give an example of how to fit device material measurements to one of the device models in \ac{AIHWKIT}, use this configuration to train a DNN with one of the specialized \ac{AIMC} training algorithms, and how to evaluate the impact of some of the device properties on the achievable accuracy. 

\subsection{Analog in-memory Training: From Device Measurements to DNN Accuracy}
\label{sec:training_fit}

Notebook {\texttt{analog\_training.ipynb}}\cite{analog_training} provides an example of how the \ac{AIHWKIT} can be used to evaluate the performance of newly characterized devices in the context of analog \ac{IMC}. The notebook starts by introducing the \RPUConfig, which is used to define many of the hardware aspects of the analog tile used in the simulation. Properties such as the tile size (which is the number of devices used in row and columns), the number of bits used by the \ac{ADC}/\ac{DAC} converters and others, as well as the specialized update algorithm and material properties, can all be defined within the \RPUConfig. 

A typical scenario in the development of new device materials includes iterations of device fabrication, characterization, and evaluation of their performance for the application under study, in this case full-scale DNN training. In particular, one wants to understand how the fabricated device performs when used in a \ac{AIMC} accelerator.
To illustrate the steps involved, we show how device measurements can be fitted to one of the \ac{AIHWKIT} device models, and then show how its impact on training accuracy can be evaluated. 

\begin{figure}[th]
    \centering
    \includegraphics[width=\textwidth]{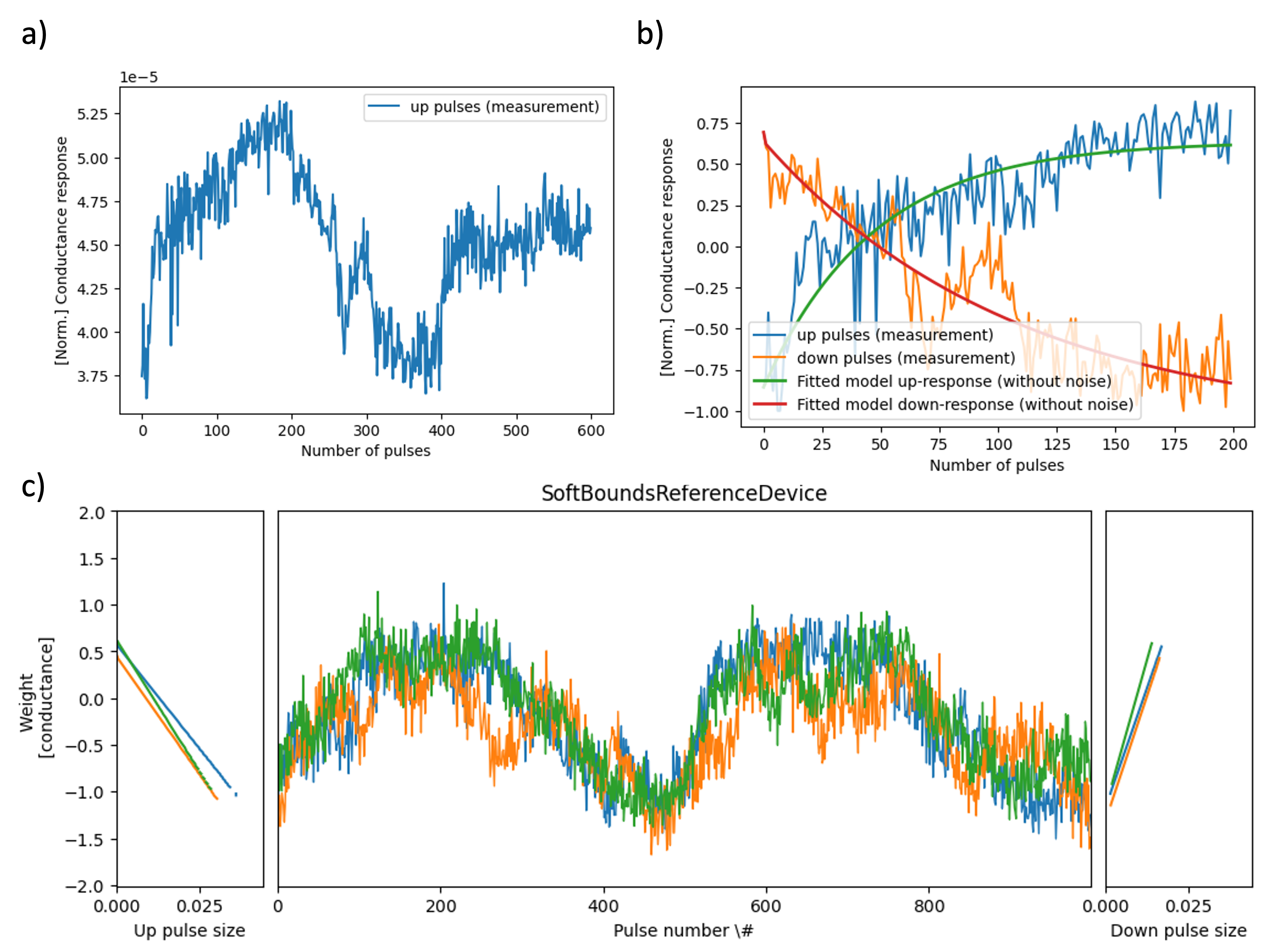}
    \caption{a) Response curve of an \ac{RRAM} memristive element. 200 pulses to increase the conductance followed by 200 pulses to decreases give the maximum and minimum conductance that the device can reach. The following 1 increase 1 decrease pulses are repeated 100 times to find the symmetry point of the device. b) The response curve and fitted model. c) Modelled response curve with noise and device-to-device variation.}
    \label{fig:dev_resp_fit}
\end{figure}

\subsubsection{Fit device measurements to a device model provided by AIHWKIT}

In \figref{dev_resp_fit}a, a typical conductance response of an \ac{RRAM} device to voltage pulse trains is shown. Here, to span the full conductance range of the device, a sequence of 200 electrical pulses is given and incrementally increases the device conductance, followed by 200 pulses which decrease the device conductance. This 200~up/200~down sequence is then followed by a 1~up/1~down pulse sequence which moves the conductance of the device to its symmetry point~\cite{gokmen2020, kim2019zeroshifting}. 

In the \ac{AIHWKIT}, there are many different device models that can be used to represent the electrical response of different devices. \ac{RRAM} electrical response is well represented through the \longvar{SoftBoundsReferenceDevice} model~(see \tabref{device-configs}), where the conductance response is gradually saturating to some level and the symmetry point can be implicitly controlled by a tunable reference device. The device models typically have many parameters, which can be fitted to the measured device characteristics. Among others, \longvar{w_max} and \longvar{w_min} represent the maximum and minimum analog weight\footnote{Note that the weight in this context refers to the \emph{analog weight} $\analog{W}$, that is the normalized conductance value that is stored in physical units in the crossbar devices (see \secref{mapping}). The full mathematical weight that is responsible for the \ac{MVM} as seen by the next layer, is given by the product of the analog weight and the digital output scales after the \ac{ADC}.} value that the device can represent, respectively (in normalized conductance units of the analog weights); \longvar{dw_min} represents the mean of the distribution of the weight change that the device can achieve at symmetry point with standard deviation of \longvar{dw_min_std}; \longvar{dw_min_dtod} specifies the device-to-device variation of the \longvar{dw_min} parameter, so that different devices can have a slightly different response curve. 

The \ac{AIHWKIT} provides a fitting utility \longvar{fit_measurements} that can be used to extract most of the needed parameters from device measurements. The notebook shows how the fitting utility can be used to automatically fit the a device model to the measured response curve shown in \figref{dev_resp_fit}~b. Since in this example the model is extracted from a single device, a device-to-device variation of 10\% is assumed.
\figref{dev_resp_fit}~c shows the fitted device response curve when noise and device-to-device variation is applied to simulate the real device characteristic. After fitting the device model, the device configuration can now be used to customize the \RPUConfig\ used in the \ac{AIMC} training.

\subsubsection{How to Specify the \RPUConfig\ for In-Memory Training}

Assuming the fitted device configuration is now given as \longvar{device_config_fit} (see notebook for an example), we can now build the \RPUConfig\ to describe the hardware and algorithmic choices of the DNN in-memory training simulation. As described in \secref{rpu-config}, the \RPUConfig\ defines many more aspects of the analog tile hardware than just the device material behaviour, for instance nonidealities in the forward and backward pass, as well as the digital periphery choices~(see \tabref{rpu-configs-fields}).  As illustrated in more detail in the notebook, we can build the following \RPUConfig\ for SGD in-memory training, where we set some (here arbitrary selected) non-default parameters:  

\begin{minted}{python}
from aihwkit.simulator.configs import (
    SingleRPUConfig, UpdateParameters, IOParameters
)
rpu_config = SingleRPUConfig(
    device=device_config_fit, 
    forward=IOParameters(out_noise=0.1),
    backward=IOParameters(out_noise=0.1),
    update=UpdateParameters(desired_bl=10),
)
\end{minted}

In this training example, in-memory training with stochastic pulses is used to train the network. Before training the DNN with this \RPUConfig\ and device setting, a DNN needs to be constructed and converted to use \emph{analog tiles} that are roughly equivalent to \ac{AIMC} crossbars. 

\subsubsection{Construct the Desired DNN and Convert to Analog}
\label{sec:model_definition}
While the \ac{AIHWKIT} provides analog layers to directly build up an analog DNN (see \tabref{analog-layers}), it is often more convenient to automatically convert a native \pytorch\ model into an analog model using the provided conversion utilities. As we show in the notebook, the native \pytorch\ \ac{DNN}, here a three layer fully connected network defined using the standard \pytorch\ syntax, is converted to an analog model by the \longvar{convert_to_analog} utility. This utility translates the layers with parameters (i.e. the three fully connected layers in this case) to be simulated with \ac{AIMC} tiles, whereas other layers, such as Sigmoid and Softmax activation function, are kept and thus assumed to be processed in digital at full precision. In the \ac{AIHWKIT}, it is generally assumed that analog signals are converted back to digital numbers after each tile operation, so that activation functions and other layers can be computed in \ac{FP}. Because \ac{AIHWKIT} is a functional simulator that aims to compute the attainable accuracy with configurable \ac{AIMC} nonidealities, and is not concerned with performance or latency estimation, the digital layers simply use native \pytorch\ code assuming floating-point precision.

\subsubsection{Train the Analog Model and Inspect the Results}
\begin{figure}
    \centering
    \includegraphics[width=0.65\textwidth]{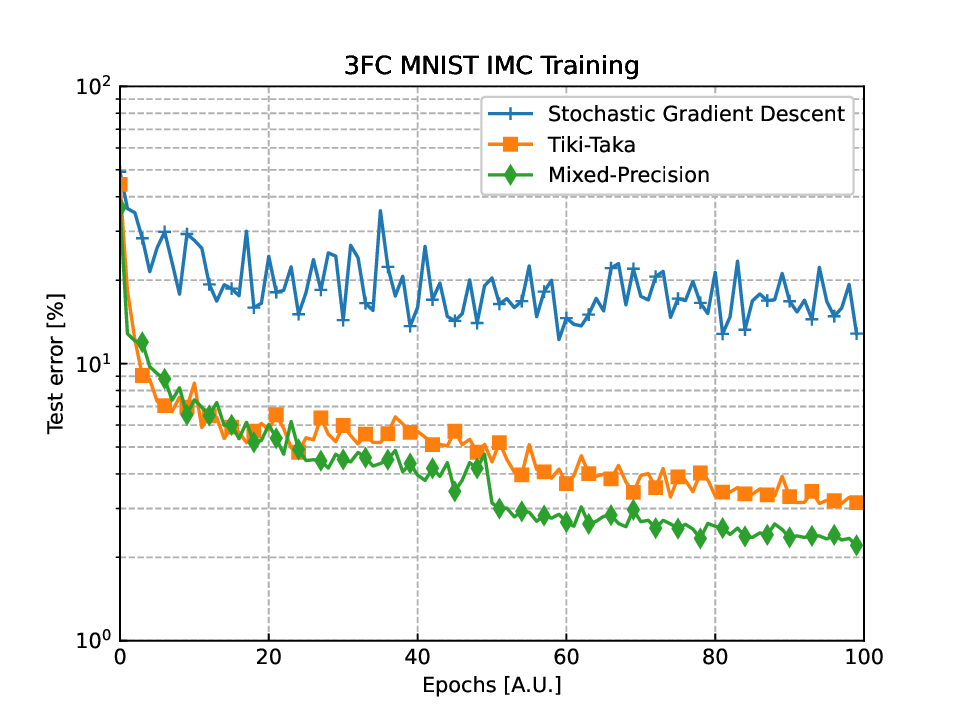}
    \caption{{Accuracy achieved by the different algorithms after 10 epochs of training. The \acf{TT} and Mixed-Precision algorithms, being specifically designed around \ac{IMC}, clearly outperform more standard \ac{SGD} algorithm}}
    \label{fig:DNNTrain}
\end{figure}

The constructed network is trained on the MNIST dataset for $100$ epochs with a batch size of 10 and a learning rate starting at 0.1 which is further decayed at the 50th and 80th epoch by a factor of 10 each time. \figref{DNNTrain} shows the performance achieved during training and compares the performance achieved by different analog in-memory training algorithms. The naive \ac{SGD} performs quite poorly, which is mainly due to the limited number of states and the asymmetry in up versus down response, as standard \ac{SGD} requires very symmetrical update characteristics~(see \cite{gokmen2016pulsed} or \cite{raschDT2020} for device specifications of \ac{SGD}). Therefore, this example shows the need for innovation not only at device level, to limit the device nonidealities and obtain a better response curve, but also at algorithmic level, to relax some of the requirements on the \ac{AIMC} device.

\subsubsection{Selecting different in-memory training optimizers}
In the above, we only used the standard in-memory \ac{SGD} training using stochastic pulse trains. The choice of other update algorithms is done by configuring the \longvar{device} field with the appropriate compound (see \tabref{compounds}). To make the building of the \RPUConfig\ more convenient, the \longvar{build_config} tool exists:

\begin{minted}{python}
from aihwkit.simulator.configs import build_config
algorithm = 'ttv2'  # one of tiki-taka, ttv2, c-ttv2, mp, sgd, agad
rpu_config = build_config(algorithm, device=device_config_fit)
\end{minted}

The notebook shows some more algorithmic choices. In this regard, the \acf{TT} algorithm~\cite{Gokmen2017} is specifically designed for training neural network with non-ideal devices. Both \ac{SGD} and \ac{TT} use error backpropagation to train the network, however the \ac{TT} algorithm replaces each weight matrix $W$ with two matrices, referred as $\analog{A}$ and $\analog{W}$. The gradients are accumulated directly onto $\analog{A}$ (using pulse coincidence) for a certain number of updates before being transferred to $\analog{W}$. \figref{DNNTrain} shows the improved performance that the \ac{TT} algorithm achieves. In contrast to the \ac{TT} algorithm, the \ac{MP} optimizer~\cite{nandakumar2020} uses digital compute for the update of gradient accumulator matrix instead of gradient accumulation in-memory. The accumulated gradient matrix $M$ is kept and computed in floating-point digital precision and then used to update the (analog) weight matrix. Given that gradients are accumulated flawlessly (but without the benefits of in-memory acceleration of the update pass), we expect that the accuracy is improved for \ac{MP}. 

\subsection{Optimize Hyperparameters of the Analog Optimizer}
As common in \ac{SGD} training, algorithmic hyperparameters such as the learning rate need to be tuned for a given AI workload. Similarly, the specialized analog optimizers come with a number of additional hyperparameters that often need to be tuned to obtain the best training result for a given device material configuration. Examples of such algorithmic hyperparameters for e.g. the \ac{cTTv2} algorithm~\cite{rasch2023fast} are specified in the "compound"-level of the device field (see also \tabref{compounds}). For instance, the parameters \longvar{auto_granularity} and \longvar{in_chop_prob}, that govern the (inverse of the) learning rate onto the $\analog{W}$ matrix and the chopper probability, respectively,  can be set with

\begin{minted}{python}
rpu_config = build_config("c-ttv2", device=device_config_fit)
rpu_config.device.auto_granularity = 2.0
rpu_config.device.in_chop_prob = 0.1
\end{minted}

Note that in this case the \longvar{device} is of a \longvar{UnitCell} type (as detailed in \tabref{compounds}). While default hyperparameter values are set to result in reasonable training behaviour, depending on the optimizer,  other hyper-parameters might need to be tuned, such as the learning rate onto the gradient accumulation matrix (\longvar{fast_lr}) or the rate of transfer reads. See \cite{rasch2023fast} for a more detailed discussion of the \ac{TT} optimizers and its variants.   

\begin{figure}
    \centering
    \includegraphics[width=0.5\textwidth]{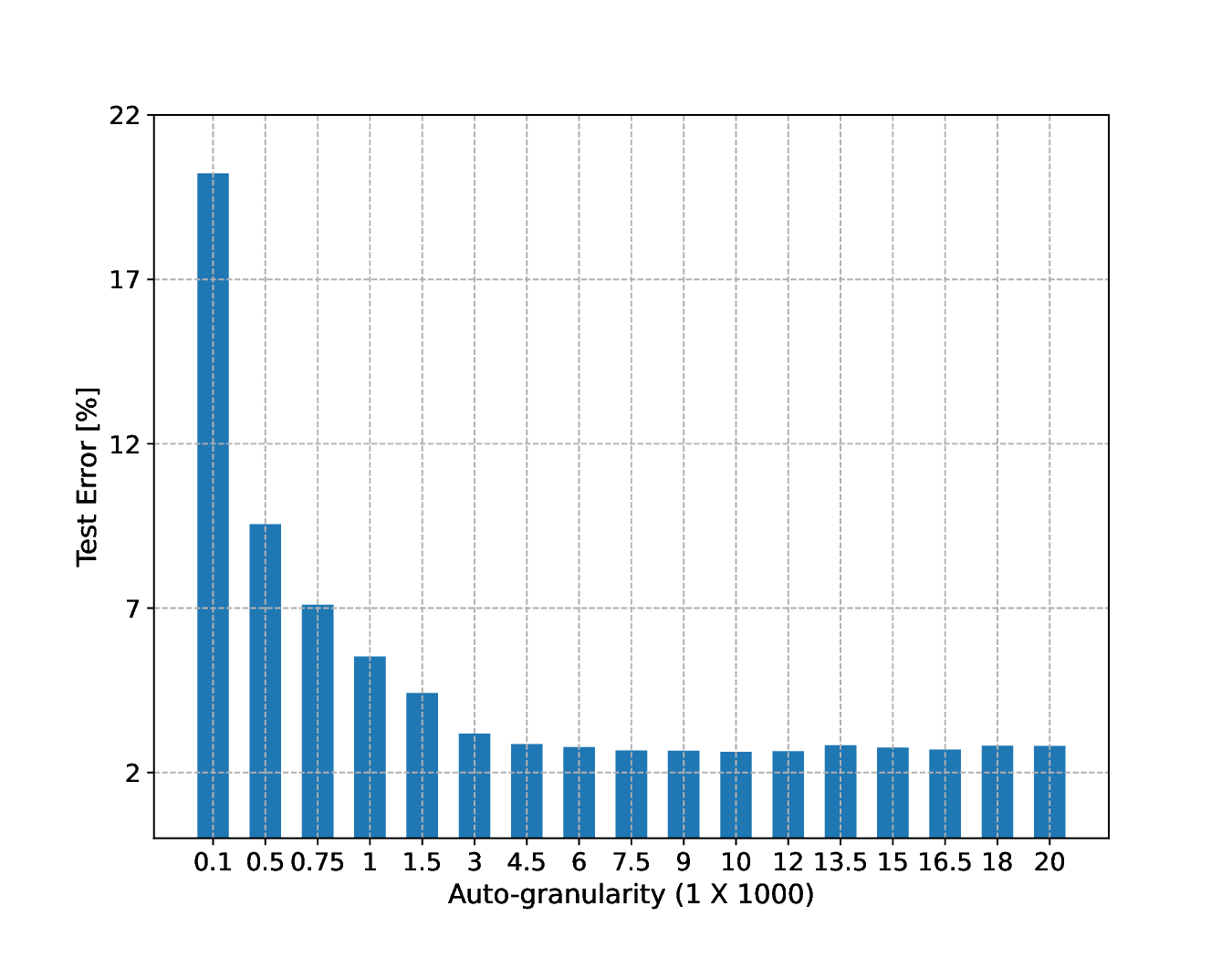}
    \caption{Validation test error (100\% - accuracy) achieved for different \longvar{auto_granularity} values when using \acf{cTTv2} in-memory training algorithm. Each data point is generated using a new \RPUConfig\ setting with adjusted \longvar{auto_granularity} value and using this new \RPUConfig\ to train the model on the train set and then tested on the separated validation set. The model used for this experiment is the model defined in 
    \secref{model_definition} and the base \RPUConfig\ is the \longvar{ReRamArrayOMPresetDevice} with \longvar{dw_min_factor} of 1.0 defined in \secref{custom_rpu_configg}.  }
    \label{fig:auto_granularity_variation}
\end{figure}

In practice, these hyperparameters need to be optimized on a separate validation data set to obtain optimal \ac{AIMC} training results for a given device model. Here, we show the effect of \longvar{auto_granularity} on the model inference error (see \figref{auto_granularity_variation}). Note that the validation test error reduces with larger \longvar{auto_granularity} values, which essentially increases the amount of noise averaging on the digital hidden matrix of the \ac{cTTv2} algorithm used here.  

\subsection{Device Parameter Variation to Obtain Device Specifications}
Apart from directly evaluating the impact of a measured device on \ac{DNN} accuracy, often one is interested in the impact of certain selected device properties on accuracy. That could be used in particular for material innovations and process developments, but also for algorithmic improvements. For instance, what is the largest device-to-device variation one can accommodate for successful training with a given analog optimizer? If known, useful design targets can be set for device material improvements. In the \ac{AIHWKIT}, all device properties can be easily varied so that such targets can be conveniently obtained as illustrated in this section. 

Although many device models are available in \ac{AIHWKIT} (see \tabref{device-configs} and \figref{device-plots}) which have a various number of parameters, some properties are common, such as \longvar{dw_min} that governs the amount of conductance change induced by a single pulse. A typical characteristics of the device is the (average) number of states, which is defined by dividing the average conductance range by \longvar{dw_min}. In most device models, the number of states and conductance ranges are thus configured by the parameters \longvar{dw_min}, \longvar{w_max}, \longvar{w_min} and their standard deviations (across pulses and across devices). Here, we take a closer look at the effect of some of these common material specifications and the algorithmic hyperparameters the attainable accuracy of the trained model in comparison to standard \ac{SGD} floating point training. As an small example \ac{DNN}, we again use the \acf{3FC} model defined and trained in \secref{model_definition} on the MNIST dataset. 

\subsubsection{Setting the \RPUConfig\ with Custom Device Parameters}
\label{sec:custom_rpu_configg}
 The basic \RPUConfig\ used in this section is based on the material specification obtained from \ac{RRAM} arrays~\cite{gong2022deep}. This device is already available in the preset library of the \ac{AIHWKIT} and is named \longvar{ReRamArrayOMPresetDevice}. Here, we investigate the impact of different values of various parameters describing the mean response and variability, such as \longvar{dw_min}, \longvar{dw_min_std}, \longvar{write_noise}, \longvar{w_max}, \longvar{w_max_std}, \longvar{w_min}, \longvar{w_min_std} etc.

 We first train the \ac{3FC} model with the given device material configuration using the specialized analog in-memory \ac{cTTv2} algorithm~(see also \tabref{compounds}). The impact of changes in the device-to-device variation, the \longvar{dw_min}, which determines number of states in the device, are investigated. To achieve this, a multiplicative factor is introduced for each parameter of interest. For example, a \longvar{dw_min} factor is introduced for changing the \longvar{dw_min} value.  This could be achieved as follows:
\begin{minted}{python}
from copy import deepcopy
from aihwkit.simulator.configs import build_config
from aihwkit.simulator.presets import ReRamArrayIEDM2022PresetDevice
def get_rpu_config(device_config, dw_min_factor = 1.0):
    device = deepcopy(device_config)
    device.dw_min *= dw_min_factor
    return build_config("c-ttv2", device=device)
# example of increasing the number of states by 2
rpu_config = get_rpu_config(ReRamIEDM2022PresetDevice(), 0.5)
\end{minted}
The resulting \RPUConfig\ due to change in a parameter factor is then used to train the model until convergence and the evaluation accuracy on the test set is stored. Similar functions can be defined for other variations (see notebook~\cite{analog_training} for more examples)

\subsubsection{Impact of Number of States of the Device Material}
\label{sec:dw_device_spec}
The number of states of the device material has a large impact on the in-memory training quality as shown in the following. The number of states of the devices is here defined by the ratio of the average conductance range and the expected response magnitude at (algorithmically) zero. Note that the device conductance in the \ac{AIHWKIT} is usually modelled in dimensionless parameters, assuming that the conductance is directly proportional to the analog weight value. All parameters of the \ac{AIMC} \ac{MVM} are in relation to this normalized unit. Thus, typically the \longvar{w_min} and \longvar{w_max} values of the device are fixed to $-1$ and $1$ respectively. Digital output scales can be used to set the analog weights initially as described in \secref{mapping}). For training, we use default torch weight initialization ranges and set the weight scaling omega to $0.3$ so that the analog weights are guaranteed to be filled with uniform numbers from $-0.3,\ldots, 0.3$ (independent on the layer size~\cite{raschDT2020}) and then fix the output scales during training. This is achieved by setting
\begin{minted}{python}
rpu_config.mapping.weight_scaling_omega = 0.3
rpu_config.mapping.learn_out_scaling = False
rpu_config.mapping.weight_scaling_columnwise = False 
\end{minted}

When we now set the \longvar{dw_min} parameter, the value is in normalized conductance units, ranging from $-1$ to $1$, so that the number of states is simply 2 divided by the value of \longvar{dw_min}.

\begin{figure}
    \centering
    \includegraphics[width=0.45\textwidth]{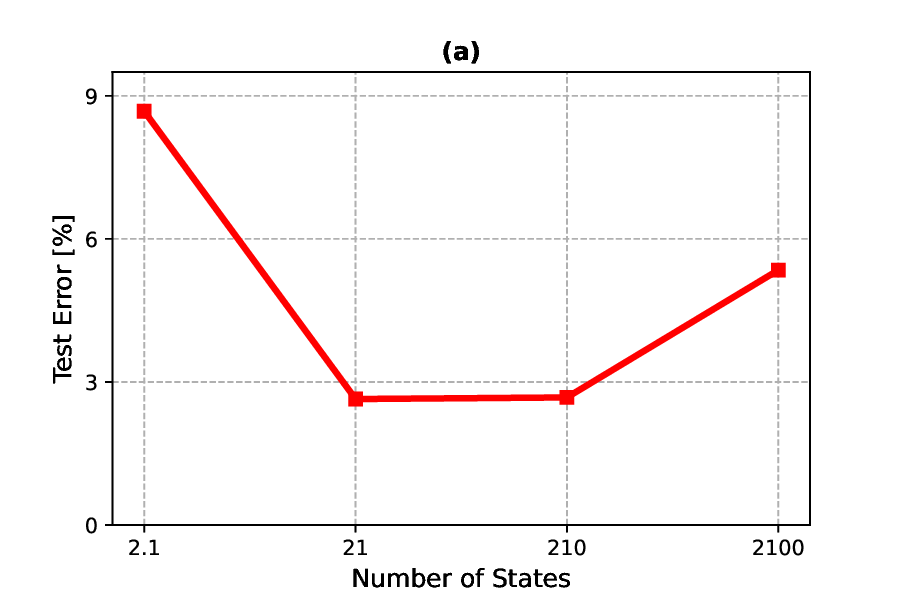}
     \includegraphics[width=0.45\textwidth]{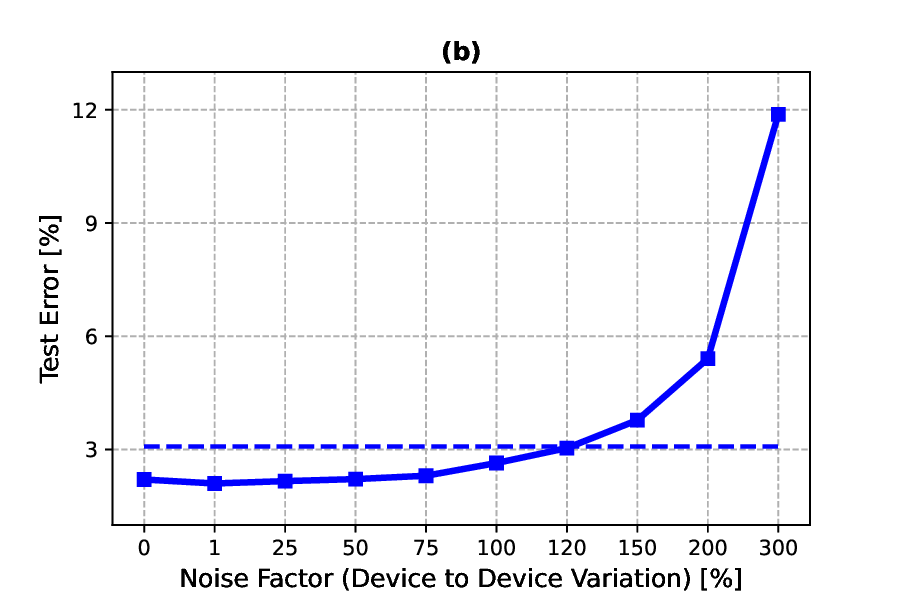}
    \caption{(a): Test error achieved by different \longvar{dw_min} factor using \ac{cTTv2} algorithm (here plotted as the number of states which is defined as the weight range divided by \longvar{dw_min}). Each data point is generated using new \RPUConfig\ obtained by varying the \longvar{dw_min} factor factor only and using new \RPUConfig\ to train the model to obtained the test error. (b): Test error achieved by the different device-to-device variation noise factor using the \ac{cTTv2} algorithm. Each data point is a the test error of a separate training run with modified \RPUConfig\ obtained by varying the device-to-device variation noise factor. The dashed line indicates 99\% of the accuracy achieved with no variation noise. This shows that about 120\% of the noise (in respect to the noise as measured for the device data at 100\%) is tolerable without significant accuracy drop.}
    \label{fig:dw_min_variation}
       \label{fig:noise_variation}
\end{figure}

\figref{dw_min_variation} (a) shows how the model test error varies with changes in the \longvar{dw_min} factor. It shows there is a range of \longvar{dw_min} values that improve the attainable accuracy, which thus means that one should optimize the device materials to match the requirements. Note, however, that too low \longvar{dw_min} factor values (higher number of states) also negatively impact the accuracy.  

\subsubsection{Impact of Device-to-Device Variations on Accuracy}
\label{sec:noise_device_spec}
Similarly, the impact of device-to-device variations can be estimated. For that, a similar parameter factor, called the noise factor and measured in percentages, is introduced as a multiplicative factor to control write noise and the various standard deviation in the material specification in relation to the baseline. A noise factor value of zero means that there is no device-to-device variation.

\figref{noise_variation} (b) shows how the device-to-device variation noise factor influences the inference performance. The validation error generally increases when increasing the noise factor, suggesting that variations negatively impact the in-memory training.  However, the rate of change is very small for noise factors between $1\%$ and $70\%$,  compared to the rate of change when the noise factor becomes greater than about $70\%$. Hence, reducing the noise variation to 70\% of the baseline might significantly improve the in-memory training performance and thus could be a helpful target for next device material designs. 

\clearpage
\section{Analog AI Cloud Composer}\label{sec:composer}

In the following, we describe the \ac{AAICC} platform, a cloud offering that provides the benefits of using the \ac{AIHWKIT} simulation platform in a fully managed cloud setting. The Analog Composer is introducing for the first time \emph{Analog AI as a service} or in short, \textit{AAaaS}. The cloud composer can be freely accessed at \url{https://aihw-composer.draco.res.ibm.com}.

We first describe the architecture of the cloud composer, and then the various services it provides including inference, training, and hardware access. We then present future features and directions. 

\subsection{Composer Design and Architecture}
The \ac{AAICC} is a novel approach to \ac{AIMC} that leverages the \ac{AIHWKIT} simulation platform to allow a seamless no-code interactive cloud experience. With access to the open-source \ac{AIHWKIT} libraries and an easy-to-use interface, it provides a platform for researchers, hardware-engineers, developers, and enthusiasts to explore, experiment, simulate, and create Analog AI neural networks and tune various analog devices to create accurate and sustainable AI models. This platform also serves as an educational tool to democratize \ac{IMC} and introduce its key concepts. 

The \ac{AAICC} adopts a modern distributed architecture based on IBM Cloud services and guidelines. The user input is limited to data (not code) with strong control and validations during the lifecycle of the application and the input data. 
The design maintains separation of concerns and responsibilities between the various components. Tracking, monitoring, and auditing services are enforced to ensure the security and compliance according to IBM Cloud security standards. 

\begin{figure}
    \centering
    \includegraphics[width=0.95\textwidth]{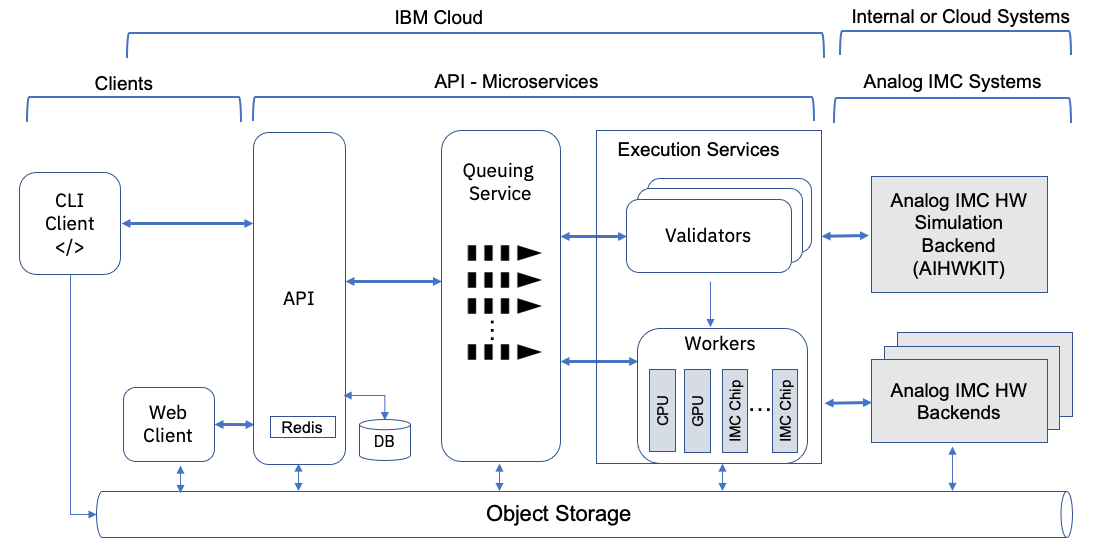}
    \caption{\acf{AAICC} Architecture}
    \label{fig:composer_architecture}
\end{figure}

The architecture of the \ac{AAICC} can be divided into five main components as illustrated in \figref{composer_architecture}:

\paragraph{The Front-end Client Component} 
This component provides an entry point for clients to interact with the composer application. Two scenarios are supported. The user can interact with the composer through a web application or through the command-line interface. Through this component, the user defines a training or inference experiment that can run on the \ac{AIHWKIT} simulator. 

\paragraph{The API} 
The API component is an HTTP microservice that provides the endpoints that are used by the web application and the backend python libraries. The API provides user authentication, database access, setup of the queuing system, job process flow setup, and collection of various statistics.

\paragraph{The Backend Execution Services}
These services are responsible for executing all the training and inference jobs that are submitted by the end users. There are two sub-components in the execution services: the validator and the workers. The validator service ensures that all training and inference jobs that are submitted are composed correctly and adhere to the specifications of the \ac{AIHWKIT} for defining the neural network, the various training or inference parameters, and the supported hardware configurations. For example, it validates that the analog layers and the \RPUConfig\ are correctly defined. The workers are responsible for executing the submitted experiments and for sending the results to the front end component for web rendering. Various worker instances are implemented depending on the backend infrastructure that will be used to execute the experiment. We have implemented three workers. The GPU worker provides GPU acceleration for running the \ac{AIHWKIT} training or inference experiments. The CPU worker will run the submitted experiments on a CPU backend server. The \ac{IMC} hardware worker will run the experiments on supported \ac{IMC} chips. The design is flexible as it allows to plugin more workers as more \ac{IMC} chips are implemented or different backend implementations are added. The infrastructure is also based on Kubernetes which allows automatic scaling of the resources depending on the load the application receives from the end-users.

\paragraph{The Queuing Services} 
This component provides the asynchronous-based communication backbone between the various components of the composer application. It implements various message queues for the CPU, GPU, and any future \ac{IMC} hardware backends. The CPU and GPU queues are used to route jobs to the \ac{AIHWKIT} backend simulation library and receive notifications of the results when the jobs are done. The IMC hardware queue(s) are used to route the jobs to analog \ac{IMC} chips that will be supported on the platform. Additionally, we have a validator queue that serves the communication between the validator and the execution workers. 

\paragraph{The Backend Analog IMC Systems} This component provides access to the \ac{AIHWKIT} for simulating training or inference on a variety of \ac{AIMC} hardware options. Real \ac{AIMC} chips will also be used to run inference or training on actual hardware (see \secref{fusion}).

\begin{figure}
    \centering
    \includegraphics[width=0.70\textwidth]{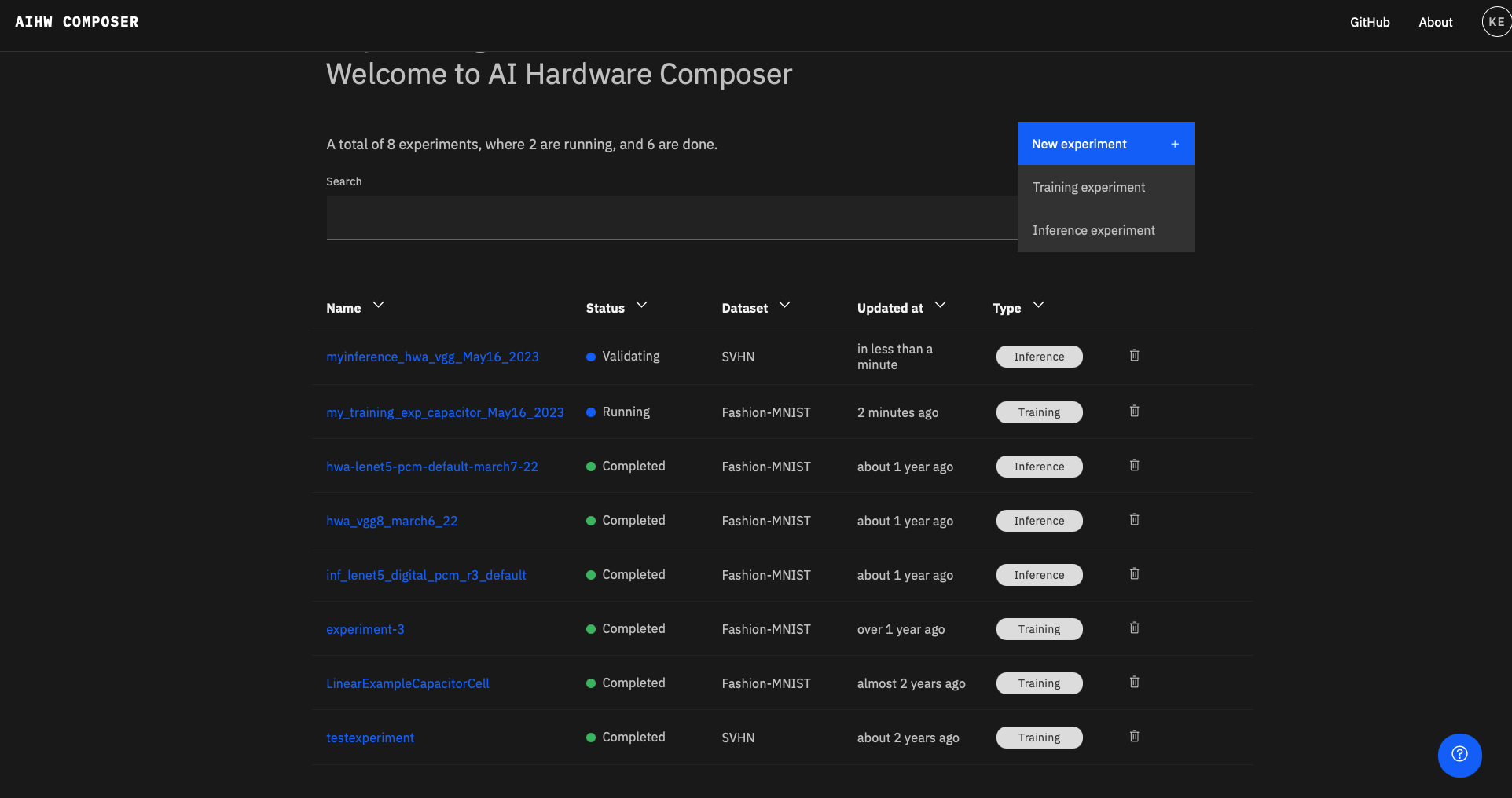}
    \caption{\acf{AAICC} Experiments Menu}
    \label{fig:composer_login}
\end{figure}

\subsection{Analog AI Training Service}
The \ac{AAICC} offers two key services: in-memory training (as explained in \secref{training}) and inference (as explained in \secref{inference}) as shown in \figref{composer_login}. In what follows we explain how these services can be used to configure, launch, and perform experiments using the \ac{AIHWKIT}. Most of the experiments are based on templates that users can choose from and customize further.

The training user experience in the \ac{AAICC} offers the end user the choice to start from an existing template or build an analog neural network from scratch. Each template provides a neural network architecture translated into analog layers or a mix of analog and digital layers, a data set, an optimizer choice, various training parameters, and an analog preset choice. We currently support templates that use the VGG8, \ac{3FC}, and LeNet \acp{DNN} for image classification tasks using various device materials and optimizer settings. This list can be easily extended as we support more neural networks and datasets.  

\begin{figure}
    \centering
    \includegraphics[width=0.70\textwidth]{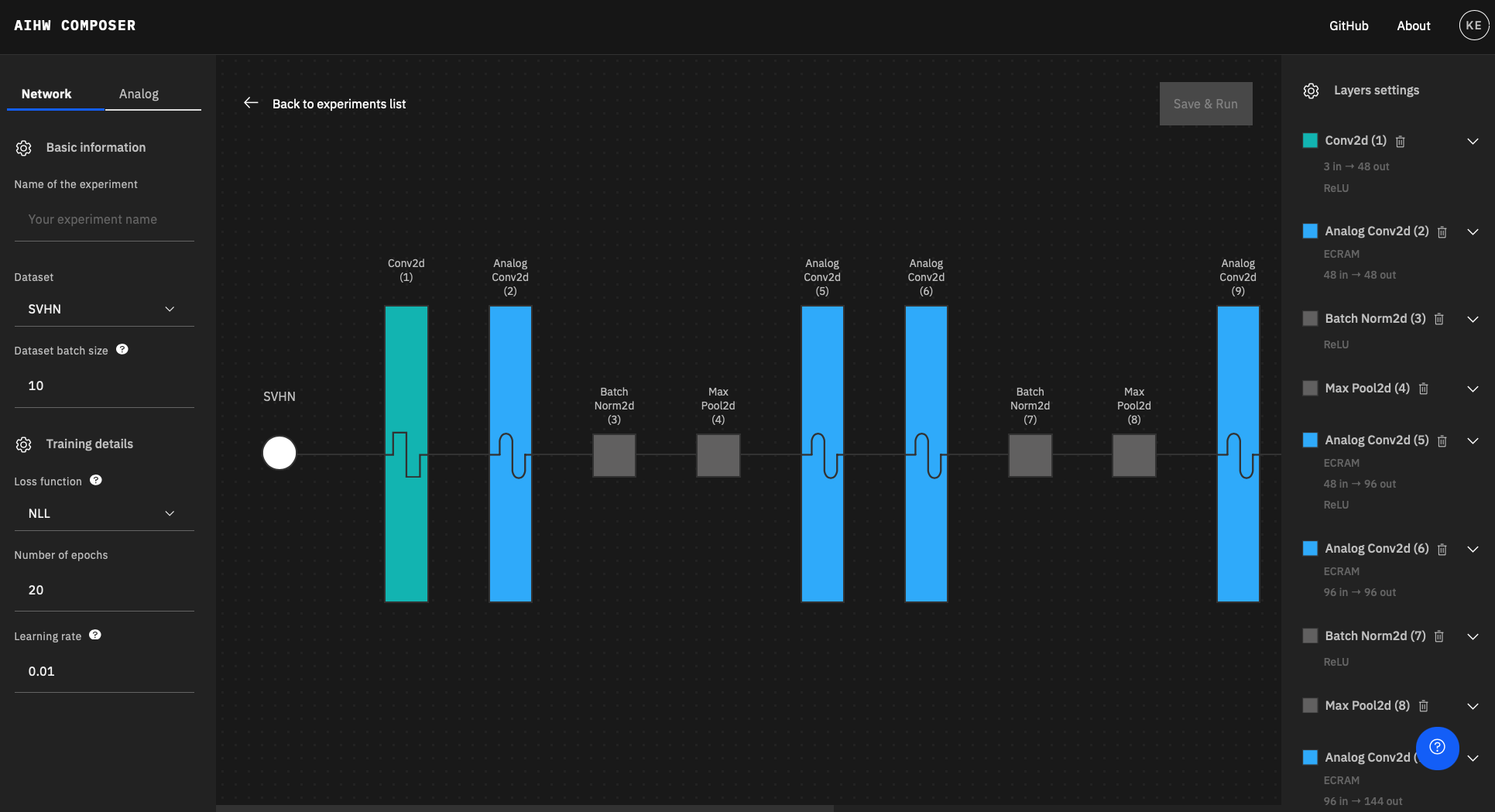}
    \caption{\acf{AAICC} Training User Interface}
    \label{fig:composer_ui}
\end{figure}

\begin{figure}
    \includegraphics[width=0.95\textwidth]{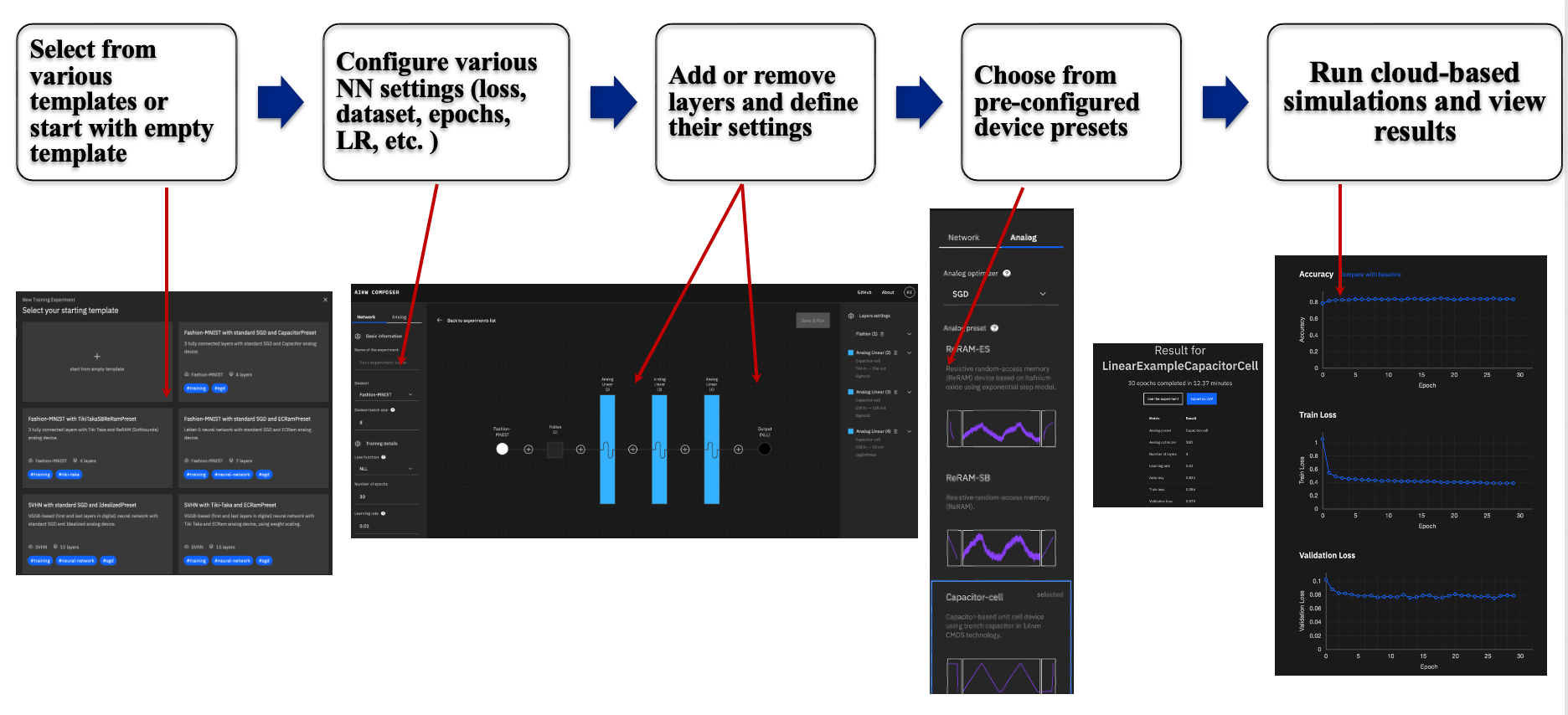}
    \caption{\acf{AAICC} In-memory Training Workflow}
    \label{fig:composer_steps}
\end{figure}

\begin{figure}[t]
    \includegraphics[width=0.70\textwidth]{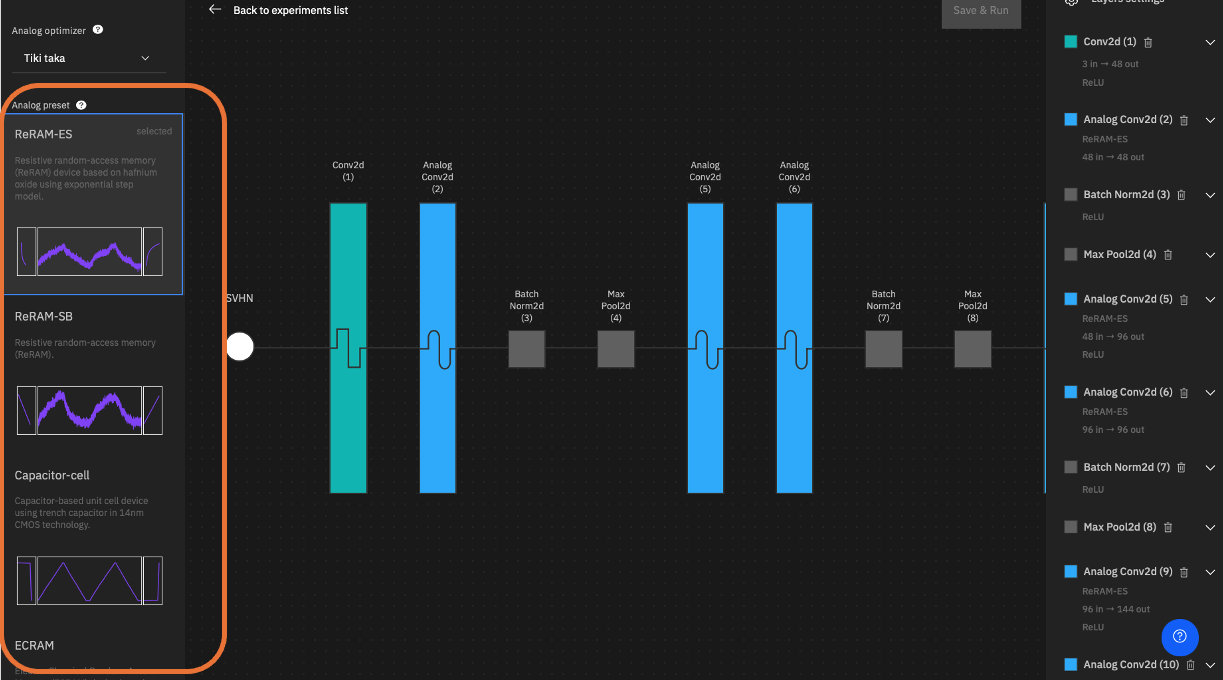}
    \caption{Apply an Analog Training Preset}
    \label{fig:composer_apply_preset}
\end{figure}

\begin{figure}
    \includegraphics[width=0.70\textwidth]{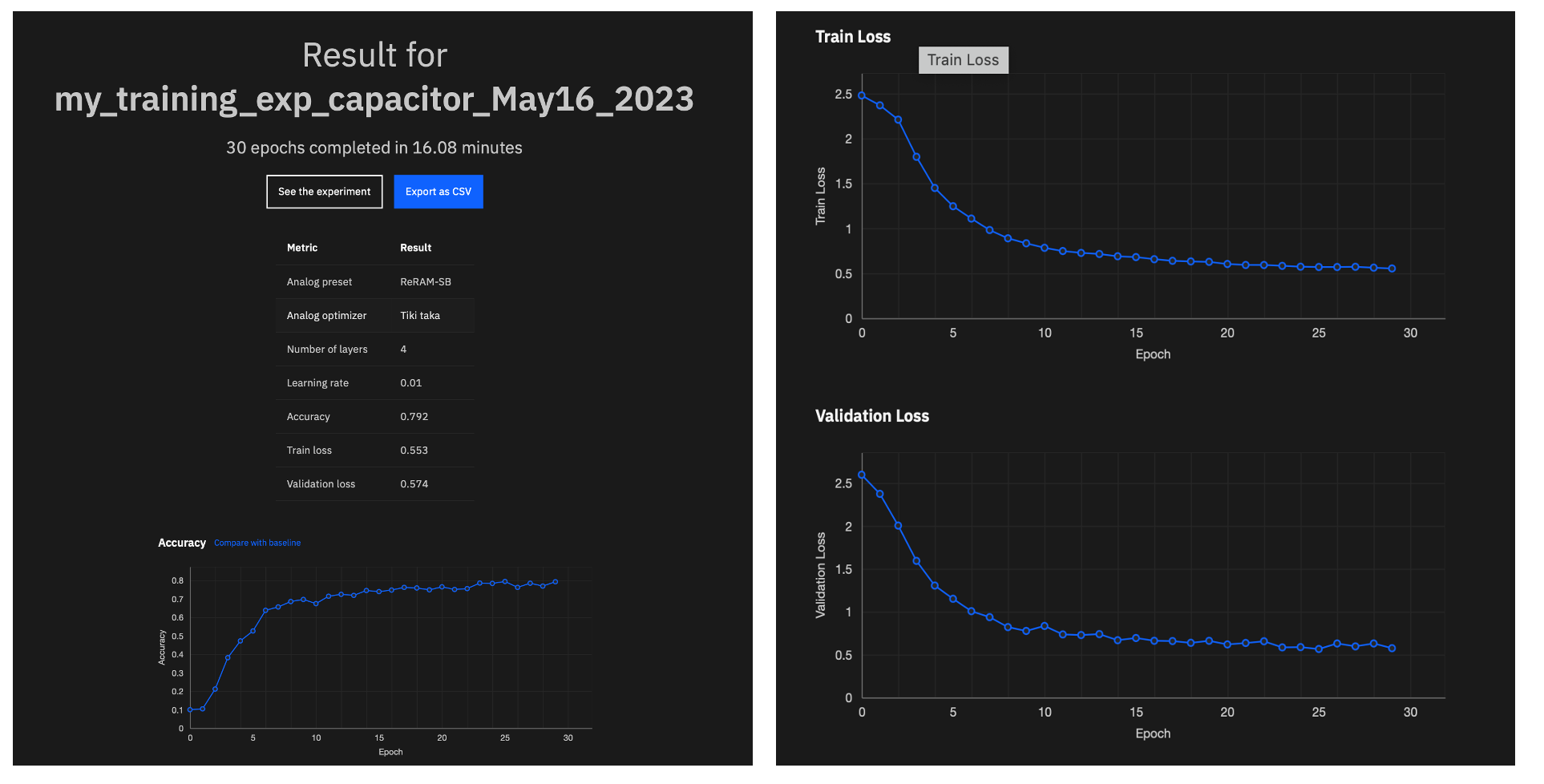}
    \caption{\acf{AAICC} Training Results Page}
    \label{fig:composer_results}
\end{figure}

The \ac{AIHWKIT} includes built-in analog presets that implement different types of devices that could be used to implement \ac{AIMC} neural network training. Many of these presets are calibrated on the measured characteristics of real hardware devices. Device non-ideal characteristics, noise, and variability are accurately simulated in all presets (see \tabref{device-configs} for a  selection of device presets). Many of these presets are readily available in \ac{AAICC} and the user can choose one of several in-memory optimizers, and thus conveniently investigate the accuracy impacts of various nonidealities and material choices on the DNN at hand.  

\figref{composer_ui} shows the composer training interface. The steps used to launch a training experiment and visualize its results are detailed below and summarized in \figref{composer_steps}:

\begin{enumerate}
    \item The user can start a new experiment or select one of the existing templates.
    \item After picking a template or choosing to compose a network from scratch, the user is then shown the composer playground interface where one can choose or configure various parameters. In the middle, the neural network layers are visualized. The left and right side of the screen provide tabs that allow the user to set training hyperparameters, analog-related configurations, or layer specific parameters. The user needs to first choose a proper name for the experiment to be created. 
    \item The next step is to input the training hyperparameters such as the batch size, the loss function, the number of epochs, and the learning rate. 
    \item The user can also add or select a layer to configure its type, size, and activation function.
    \item One of the key features of this interface is the ability to explore and apply an analog device preset as shown in \figref{composer_apply_preset}. The interface also provides useful documentation about each preset. The user can learn about the technology and device materials used in each preset and view the conductance response curve.
    \item Once the user defines all training and analog related parameters, a training experiment can be launched on the cloud by clicking on the save and run button to launch. The jobs will be accelerated by GPUs in the cloud. The experiment is validated first by the back-end to ensure correctness of the user-provided input before invoking the \ac{AIHWKIT} to run the \ac{AIMC} training simulation. 
    \item Upon completion of the training experiment, the results page as shown in \figref{composer_results} summarizes the key training job parameters that were used such as the analog preset and the analog optimizer algorithm and plots the trained model's accuracy, validation loss and training loss per epoch.  
\end{enumerate}


\subsection{Analog AI Inference Service}
\label{sec:comopser-inference}

Similar to the training service, the \ac{AAICC} inference service provides a template-based interactive no-code user experiences that allows creating analog inference experiments and launching them in the cloud. \figref{composer_inf} illustrates the high-level workflow used which is detailed below:
\begin{figure}
    \includegraphics[width=0.95\textwidth]{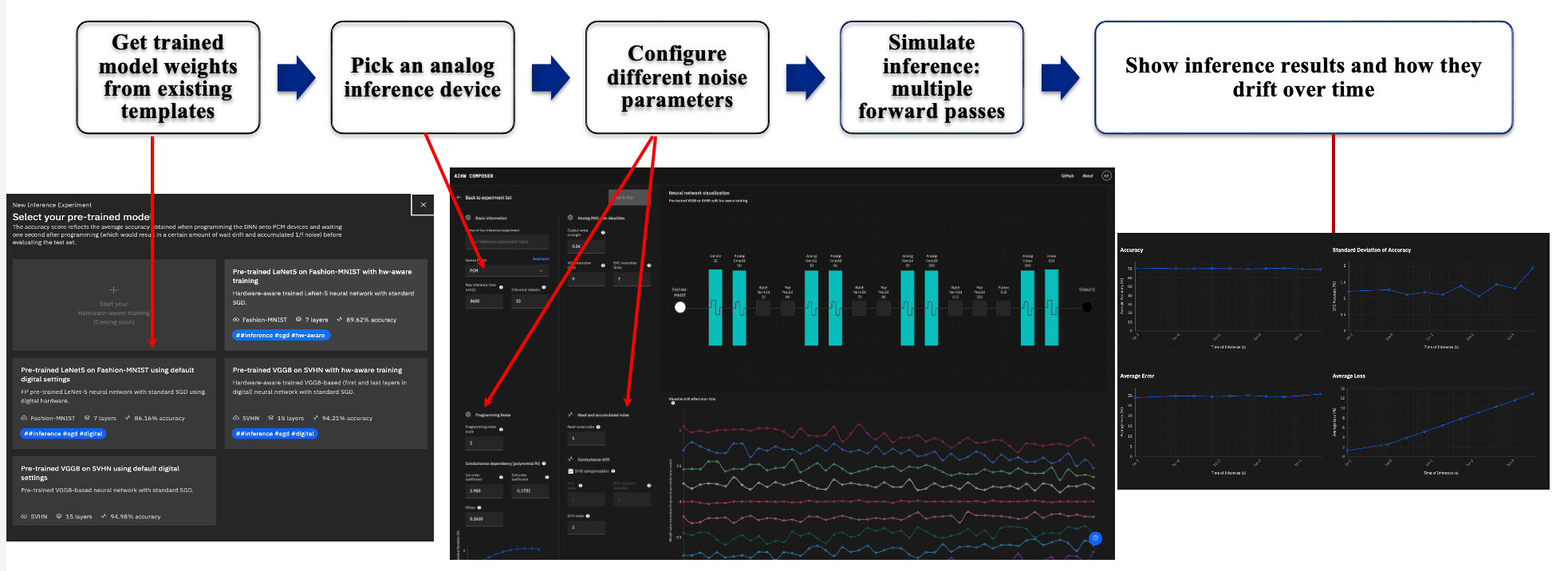}
    \caption{\acf{AAICC} Inference Workflow}
    \label{fig:composer_inf}
\end{figure}

\begin{figure}
    \includegraphics[width=0.95\textwidth]{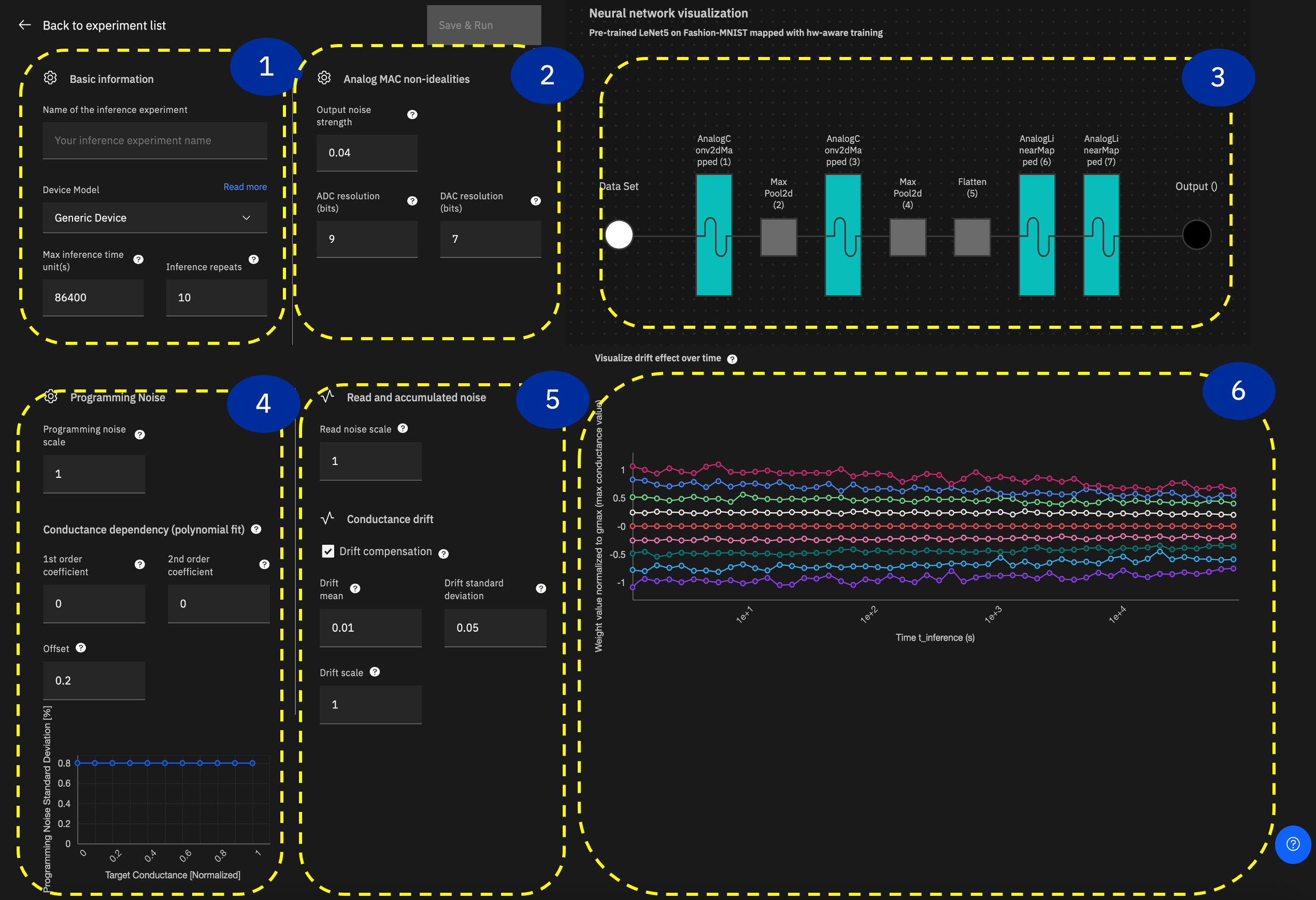}
    \caption{Design of the \acf{AAICC} Inference User Interface}
    \label{fig:composer_inf_ui}
\end{figure}

\begin{enumerate}
    \item First, the user can pick one of the pre-trained model templates. We provide models that are either hardware-aware trained or trained in digital hardware such as GPUs. There are a number of available pre-trained models and their characteristics (using VGG8 and LeNet DNNs for image classifcation with a combination of digital and analog layers). Future work will enable hardware-aware training directly from the composer interface that can feed into this interface.
    \item An \ac{AIMC} inference device needs to be chosen. We provide two choices: a \ac{PCM} abstract device or a state-independent generic device. The \ac{PCM} model (\longvar{PCMLikeNoiseModel}) is described in detail in \secref{noise-inference}. 
    \item The next step is to configure different noise parameters and drift strengths. Different \ac{MVM} nonidealities sources can be tuned to study their effect on the accuracy as shown in \figref{composer_inf_ui}. These nonideality settings correspond to the \RPUConfig\ choices for inference (see \secref{inference} and \tabref{mvm-nonidealities} for details). 

    \item Depending on the configured parameters, the inference service provides an interactive graph that visualizes the drift effect over time of the hardware device that is simulated (\ac{PCM} or generic device). As shown in \figref{composer_inf_ui}, the graph shows how the weights are drifting over time for different weight values after they have been programmed on the device.  
    \item The inference simulation using \ac{AIHWKIT} can then be launched as a job in the cloud. The user can visualize the results of the inference including model accuracy and drift effects over time.
\end{enumerate}


\begin{figure}
    \includegraphics[width=0.95\textwidth]{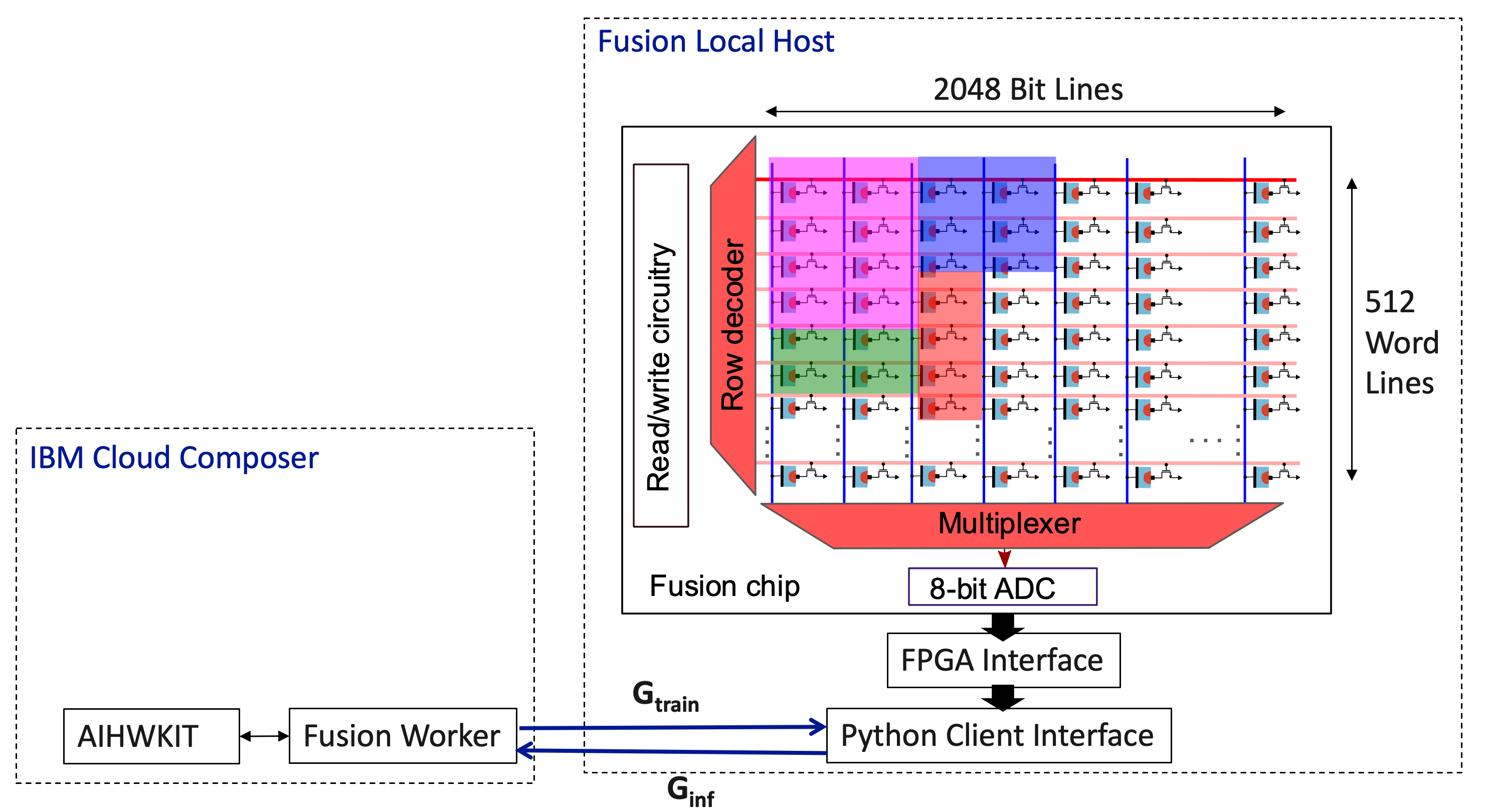}
    \caption{Access to the Fusion Chip through the \acf{AAICC}}
    \label{fig:fusion-chip}
\end{figure}

\subsection{Access to Analog IMC Hardware}
\label{sec:fusion}
\begin{figure}
    \includegraphics[width=0.95\textwidth]{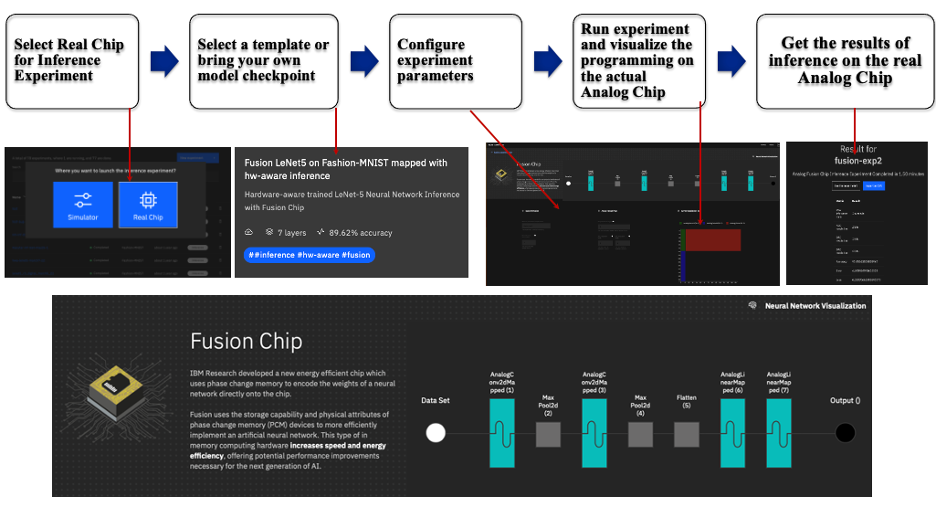}
    \caption{Workflow of Accessing the Analog Inference \ac{PCM}-based Fusion Chip}
    \label{fig:fusion-chip-workflow}
\end{figure}

In addition to the inference and training simulations using the \ac{AIHWKIT}, the composer application provides a framework for accessing real IBM \ac{IMC} chips as they become available. The first \ac{IMC} chip that we will expose is the Fusion \ac{PCM} chip~\cite{Joshi2020}. As shown in \figref{fusion-chip}, the Fusion chip has 512 word lines (WL) and 2048 bit lines (BL). Each WL/BL address has a \ac{PCM} device and an access transistor which can be individually accessed. Hence, there are $512\times2048$ \ac{PCM} devices in total. Because the chip only stores the weights and does not perform an explicit \ac{MVM} on-chip, they can be placed at any arbitrary location on the chip independently of which layer they encode. Each \ac{PCM} device stores the absolute value of a weight in its conductance state. The sign information is stored in the python client software. 
\figref{fusion-chip} shows a high level description of how the Fusion chip interacts with the composer. Trained weights from user are converted to conductance values $G_\text{train}$ and then sent to the python client running on a local host to program the weights on \ac{PCM} chip. After programming, conductance values are read from the \ac{PCM} chip through the local python client and thus provides an accurate measurement of the programmed weight and its deviation from the target conductance. The hardware conductance measurements are sent to the \ac{AIHWKIT} running on IBM cloud which will then perform inference on them and simulate the additional \ac{MVM} nonidealities shown in \tabref{mvm-nonidealities}. Inference results are displayed on the UI or can be retrieved via a command line interface. \figref{fusion-chip-workflow} shows a preview of the \ac{AAICC} user experience that allows accessing our first analog inference Fusion chip. This capability is still in beta version and under active development.

\subsection{Road-map and Future Directions}

\begin{figure}
    \includegraphics[width=0.95\textwidth]{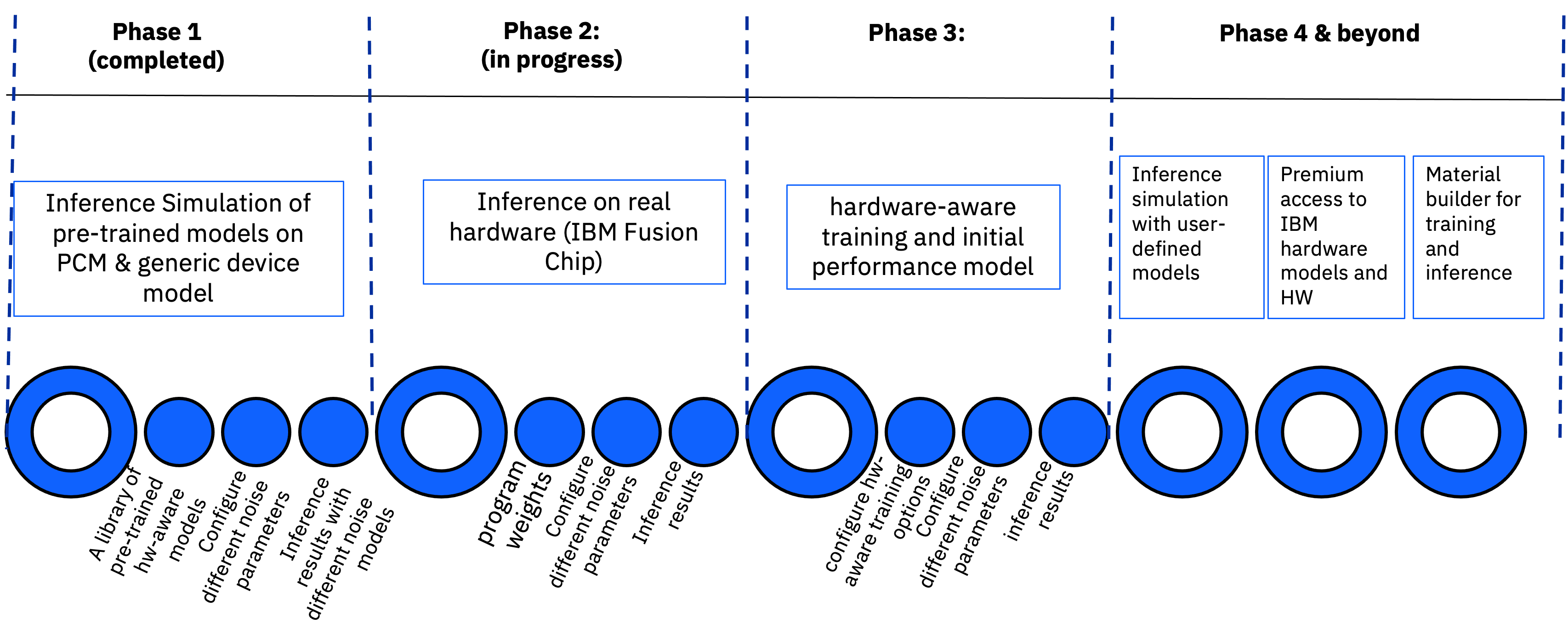}
    \caption{\ac{AAICC} Application Roadmap}
    \label{fig:composer-roadmap}
\end{figure}

A cloud no-code interactive experience has been developed to provide a platform for cloud-based experiments, access to IBM Research hardware technology and the creation of a vibrant ecosystem around Analog AI.
\ac{AIHWKIT} can be used online though a web-based front-end \ac{AAICC}. The composer provides a set of templates and a no-code experience to introduce the concepts of Analog AI, configure experiments, and launch training and inference experiments on IBM public Cloud. The future road-map includes adding hardware-aware training, energy and latency performance models' estimators, access to more IBM Research premium Analog AI chips as a service, adding additional advanced capabilities such as a material builder for training and inference, and continuing to expose the latest algorithmic innovations from IBM Research to the open-source community as consumable services. \figref{composer-roadmap} summarizes our short terms and long-term plans. 

\clearpage
\section{How to Extend and Customize the AIHWKIT}
\label{sec:extension}
The \ac{AIHWKIT} has been designed to be easily customizable and modular to ease any feature extensions.  Moreover, \ac{AIHWKIT} is implemented using modern coding and open source practices, such as python code formatting guidelines, versioning, github integration for collaborative coding, and unit testing to ensure quality and back-functionality when adding new code (see \cite{aihwkit} and the online documentation for more details).   

In the following, we give a number of examples of how to extend functionality. In particular, we show how a new phenomenological inference noise model can be added, how a custom drift compensation is implemented, and how the \ac{AIMC} crossbar simulation could be enhanced.   

\subsection{Custom Phenomenological Inference Noise Model}
Phenomenological inference noise models are applied to model the long-term noise effects to the NVM device (see \secref{noise-inference}). To capture the initial programming error, as well as the long-term temporal component of the conductance changes, \ac{AIHWKIT} allows for defining inference noise models (such as \longvar{PCMLikeNoiseModel}). 

Let us assume one has a new material and matching the measurements with the provided noise models is not possible even when changing the parameters. In this case, one needs to implement a customized noise model. For that, one needs to derive a new class from the \longvar{BaseNoiseModel}, and override a number of methods that define what noise is added. First, \longvar{apply_programming_noise_to_conductance}, that applies programming noise to given conductances (in $\mu$S) and returns the programmed conductances. Second, \longvar{apply_drift_noise_to_conductance}, that applies long-term noise (e.g. drift and 1/f noise) to the programmed conductances. Lastly, \longvar{generate_drift_coefficients}, that generates the drift coefficients during the programming that will be given as input when applying the drift, if needed. 

In the following example, we implement a very simple model that just assumes a Gaussian additive programming model and constant conductance drift. The new noise model class could look like (omitting import statements, see notebook \texttt{extending\_functionality.ipynb}\cite{extension}):

\begin{minted}{python}
class SimpleNVMNoiseModel(BaseNoiseModel):
    """Very simple noise model of a new material """

    def __init__(self,  nu=0.1, prog_std=0.1, **kwargs):
        super().__init__(**kwargs)
        self.nu = nu
        self.prog_std = prog_std  # in muS 

    def apply_programming_noise_to_conductance(self, g_target: Tensor) -> Tensor:
        """Apply programming noise to a target conductance Tensor. """
        g_prog = g_target + self.prog_std * randn_like(g_target)
        g_prog.clamp_(min=0.0)  # no negative conductances allowed
        return g_prog

    def generate_drift_coefficients(self, g_target: Tensor) ->Tensor:
        """Just constant nu"""
        return tensor(self.nu)
     
    def apply_drift_noise_to_conductance(self, g_prog, nu, t_inference) -> Tensor:
        """Apply drift up to the assumed inference time"""
        t_0 = 1  # assume 1 sec as drift reference
        t = t_inference + t_0
        if t <= t_0:
            return g_prog
        return g_prog * ((t / t_0) ** (-nu))
\end{minted}

Note that before the noise model is applied for inference accuracy evaluation (see \secref{inference}), the learned target weight values are passed through a conductance converter to get a list of conductances, for which the noise model is applied (i.e. when \longvar{analog_model.drift_analog_weights(t_inference)} is called). 

To describe this process in more detail, let us first get the target analog weight values of an analog tile. The target analog weight values are the tile weight values without applying any digital output scales. We thus get these target analog weight values with (here simply for the first analog tile of a model): 
\begin{minted}{python}
analog_tile = next(analog_model.analog_tiles())
target_analog_weights, _ = analog_tile.get_weights(apply_weight_scaling=False)
\end{minted}
These target analog weight values are however still in normalized units (typically in range $-1, \ldots, 1$), having thus both negative and positive values. To get the conductances from these normalized target analog weight values,  a conductance converter is used (see \longvar{aihwkit.inference.converter.conductance}). For instance, the \longvar{SinglePairConductanceConverter} would return a list of conductance matrices, one for positive values of the analog weights (setting negative values to zero and scaling it by $g_\text{max}$ to get values in $\mu$S) and one for negative values (setting positive values to zero and scale it similarly):
\begin{minted}{python}
g_converter = SinglePairConductanceConverter(g_max=25.0)
g_tuple = g_converter.convert_to_conductances(analog_weights)
\end{minted}
This list of conductance matrices can be converted back with
\begin{minted}{python}
new_analog_weights = g_converter.convert_back_to_weights(*g_tuple)
\end{minted}
where in this case the new analog weights are simply the old, because the noise model was not yet applied. Note that the conversion from normalized target analog weight values to conductances can also be customized by adding a new conductance converter.   

However, this conductance conversion is happening internally when the programming is applied. In other words, the noise model above is applied after the conversion to conductances and then conductances are then internally converted back to normalized analog weights and applied back to each analog tile. Therefore, to use the above new noise model one can simply do, for instance, to evaluate a ResNet:
\begin{minted}{python}
rpu_config = InferenceRPUConfig(
                 noise_model=SimpleNVMNoiseModel(prog_std=0.3,
                     g_converter=SinglePairConductanceConverter(g_max=25.)))
analog_model = convert_to_analog(resnet32(), rpu_config)
analog_model.eval()
analog_model.program_analog_weights()
analog_model.drift_analog_weights(3600) # 1 hour
\end{minted}
Now the analog weights are programmed and drifted and one could evaluate the accuracy with such long-term noise sources applied to the analog weights.

\subsection{Custom Drift Compensation}
In the above, example no drift compensation was used. Drift compensations are needed for inference with materials that exhibit conductance drift and they are modular classes in the \ac{AIHWKIT} that can be easily customized.

For instance, assume that the baseline of the drift compensation should be read multiple times (instead of a single time) to improve the signal-to-noise ratio when applying the drift compensation during inference. For that, one could implement a new custom drift class that derives from the base drift compensation class. The custom drift compensation class could look like the following (all import statements are omitted for brevity, see notebook\cite{extension} for mode details):

\begin{minted}{python}
class NTimesDriftCompensation(BaseDriftCompensation):
    """Global drift compensation with multiple read-outs."""
    def __init__(self, n_times: int = 1) -> None:
        self.n_times = n_times
    def readout(self, out_tensor: Tensor) -> Tensor:  
        return clamp(torch_abs(out_tensor).mean(), min=0.0001)
    def get_readout_tensor(self, in_size: int) -> Tensor:
        """Return the read-out tensor with n-times one-hot vectors (eye)."""
        return tile(eye(in_size), [self.n_times, 1])
\end{minted}
Now this new drift compensation can be simply set when specifying the \RPUConfig, such as
\begin{minted}{python}
rpu_config = InferenceRPUConfig(
                 drift_compensation=NTimesDriftCompensation(n_times=10))
\end{minted}
This \RPUConfig\ can then be used to define an analog model and will be used for inference evaluation as described in the previous example. 

\subsection{Modifying the AIMC MVM for each Analog Tile}
    The basic \ac{AIMC} \ac{MVM} is typically part of the C++ \textsc{RPUCuda} engine for speed and thus less easily extended using python. However,  \ac{AIHWKIT} provides a separate python implementation of (some of) the \ac{AIMC} \ac{MVM} nonidealities. This analog \ac{MVM} is encapsulated in the base class \longvar{SimulatorTile}. Here we show how one could add changes to the way the analog \ac{MVM} is performed. In this example, we only show it for inference only (deriving from the inference-only tile \longvar{TorchSimulatorTile} and modfiying the forward pass), but a custom in-memory training tile can similarly be implemented by deriving from the \longvar{CustomSimulatorTile} in \longvar{aihwkit.simulator.tiles.custom} by overriding the forward, backward, or update methods.

     In a simple example, we create a new simulator tile class that modifies the forward pass of the inference evaluation. Currently, the implementation in \longvar{TorchSimulatorTile} does negative and positive inputs in one \ac{MVM} pass. Let us assume one wants to simulate two analog MVMs instead, one for positive and one for negative inputs, and add them results together in digital.     
    
    This could be simply done by defining a new \longvar{SimulatorTile} that derives from the  \longvar{TorchSimulatorTile} but overrides the forward pass accordingly. Parameters from the \RPUConfig\ can be passed during the initialization and are considered constant.  The new class could be defined as (omitting the import statements, see notebook\cite{extension} for details): 
        
\begin{minted}{python}
class TwoPassTorchSimulatorTile(TorchSimulatorTile):
    """New class where two forwards are done optionally"""
    
    def __init__(self, x_size: int, d_size: int, 
            rpu_config: "TwoPassTorchInferenceRPUConfig", bias: bool = False):
        super().__init__(x_size, d_size, rpu_config, bias)
        self._one_pass = rpu_config.one_pass    

    def forward(self, x_input: Tensor, **kwargs) -> Tensor:
        if self._one_pass:
            return super().forward(x_input, **kwargs)
        x_pos, x_neg = clamp(x_input, min=0.0), -clamp(x_input, max=0.0)
        return super().forward(x_pos, **kwargs) - super().forward(x_neg, **kwargs)
\end{minted}

Note that this new class defines a new simulator tile that modifies the way the \ac{MVM} is computed and defines a new parameters (\longvar{one_pass}). To use this tile in a DNN we need to provide a compatible \RPUConfig\ that uses this simulator tile:

\begin{minted}{python}
@dataclass
class TwoPassTorchInferenceRPUConfig(TorchInferenceRPUConfig):
    """Optionally using two forward passes for negative and positive inputs"""
    simulator_tile_class: ClassVar[Type] = TwoPassTorchSimulatorTile
    one_pass: bool = True
    """Optionally turn on the two passes"""
\end{minted}
Now we can simple use this new \RPUConfig\ for model conversion, e.g. :
\begin{minted}{python}
rpu_config = TwoPassTorchInferenceRPUConfig(one_pass=False)
analog_model = convert_to_analog(resnet32, rpu_config)
\end{minted}
The analog model will now use the new simulator tile. 

Similarly other aspects of the \ac{AIMC} compute can be extended by an analogous approach. For instance, one could add a new peripheral (digital) computation,  which would then require to override methods of the  \longvar{AnalogTile} or \longvar{InferenceTile}, that encapsulate the full tile operations on a higher level (that is analog MVM simulations in the lower-level \longvar{SimulatorTile} and also digital periphery, such as output scaling). 

If users decide to implement custom functionality, we highly encourage to share the new code with the community. Integrating the new addition to the open source is as easy as raising a new pull request on the \ac{AIHWKIT} github.       

\clearpage
\section{Outlook}\label{sec:outlook}
Having described in detail the functionality of \ac{AIHWKIT} and how to customize it, we would like to briefly highlight a few possible research directions that could be pursued with the toolkit in this last section. The primary use-case for \ac{AIHWKIT} is, of course, the exploration of device-level parameter specifications for inference and training, which has already been the subject of several publications\cite{zhenming2022,lee2022impact,li2023impact, rasch2023hardware}. In addition, novel analog optimizers for on-chip training could be implemented and tested to demonstrate improvement over the existing ones on a wide range of device parameters \cite{rasch2023fast}. For inference, a noteworthy addition to \ac{AIHWKIT} could be to implement the auxiliary digital operations for affine scaling, batch normalization, and activation functions with low-precision arithmetic to study the digital precision requirements on a wide range of networks. Another interesting direction would be to implement input and weight bit slicing\cite{legallo2022}, and evaluate the impact of those schemes for inference and training. While (almost) arbitrary pre-trained models can already be converted by \ac{AIHWKIT} and custom trained, it would still be worthwhile in the future to make (\ac{HWA}) training compatible with other training pipeline libraries, such as DeepSpeed\cite{deepspeed}, HuggingFace\cite{huggingface}, or Fairseq\cite{fairseq}, in order to conveniently re-use preexisting code using these pipelines. Finally, extending \ac{AIHWKIT} to generate approximate power and latency estimates, using some fairly generic assumptions on the hardware architecture being modeled, would be desired to explore optimal \ac{AIMC} design approaches using neural architecture search\cite{benmeziane2023analognas}.

We hope that this tutorial will make the barrier of entry more accessible for new users to adopt \ac{AIHWKIT} to simulate inference and training of \acp{DNN} with \ac{AIMC}. \ac{AIHWKIT} not only provides accurate hardware-calibrated models of \ac{AIMC} devices and the main peripheral circuit nonidealities present in a \ac{AIMC} chip, it is also continuously being maintained by a team of developers who are actively fixing issues and adding new features. Therefore, user-made contributions to \ac{AIHWKIT} will be integrated into a well-maintained toolkit and will benefit from being further improved as the toolkit develops, instead of getting "lost" into a private repository that would involve too much overhead from the user to be maintained properly. For this reason, we strongly encourage users working in the \ac{AIMC} field to adopt actively maintained toolkits, such as \ac{AIHWKIT}, and make the effort to integrate their contributions to them. Otherwise, contributions put in individual repositories will likely get abandoned and just add up to the excessive tool fragmentation that already prevails. It is only with active contributions from the community, and by bringing all those contributions together into a single tool, that \ac{AIMC} can eventually become commercially successful and lead to a new era of efficient and sustainable non-von Neumann accelerators.


%
%

%

\section*{Supplementary Material}

The four Jupyter Notebooks that accompany the paper are accessible at \url{url}. 

\begin{acknowledgments}
We thank Geoffrey Burr, An Chen, Andrea Fasoli, Pritish Narayanan, Tayfun Gokmen, Takashi Ando, John Rozen, Irem Boybat, Athanasios Vasilopoulos, Hadjer Benmeziane and Ghazi Sarwat Syed for technical input on \ac{AIHWKIT}; Kim Tran, Kurtis Ruby, Borja Godoy, Jordan Murray, Todd Deshane, Diego Moreda and Kevin Johnson for developing the Analog AI Cloud Composer; and Jeff Burns for managerial support. This work was supported by the IBM Research AI Hardware Center. This work has also received funding from the European Union's Horizon Europe research and innovation program under Grant Agreement No 101046878, and was supported by the Swiss State Secretariat for Education, Research and Innovation (SERI) under contract number 22.00029.We thank the computational support from AiMOS, an AI supercomputer made available by the IBM Research AI Hardware Center and Rensselaer Polytechnic Institute's Center for Computational Innovations (CCI). 
\end{acknowledgments}

\section*{Data Availability}

The data that support the findings of this study are openly available in the \ac{AIHWKIT} repository at \url{https://github.com/IBM/aihwkit}\cite{aihwkit_repo}.

\bibliography{papers_mem}

\end{document}